\documentclass[12pt]{article}           
\usepackage{hyperref}
\usepackage[pdftex]{graphicx}
\usepackage{amsmath,amssymb}
\usepackage{mathabx}
\usepackage{color}
\usepackage{cite}

\newcommand{\be}{\begin{equation}}
\newcommand{\ee}{\end{equation}}
\newcommand{\bea}{\begin{eqnarray}}
\newcommand{\eea}{\end{eqnarray}}

\def\p{\partial}

\newcommand{\tr}{\mathop{\rm Tr}}

\def\ket#1{| #1 \rangle}
\def\bra#1{\langle  #1 |}

\newcommand{\cA}{{\mathcal{A}}}

\newcommand{\cC}{{\mathcal{C}}}
\newcommand{\cD}{{\mathcal{D}}}
\newcommand{\cE}{{\mathcal{E}}}

\newcommand{\cO}{{\mathcal{O}}}

\newcommand{\bt}{{\mathbf{t}}}
\newcommand{\bx}{{\mathbf{x}}}
\newcommand{\br}{{\mathbf{r}}}
\DeclareMathOperator{\csch}{csch}
\DeclareMathOperator{\arctanh}{arctanh}
\DeclareMathOperator{\arcsinh}{arcsinh}

\begin{document}

\begin{titlepage}

\vspace*{1cm}
\begin{center} \Large \bf Entanglement Wedge Reconstruction\\
and Entanglement of Purification
\end{center}

\begin{center}
Ricardo Esp\'indola$^{\ast}$,
Alberto G\"uijosa$^{\ast}$,
and Juan F.~Pedraza$^{\diamond}$

\vspace{0.2cm}
$^{\ast}\,$Departamento de F\'{\i}sica de Altas Energ\'{\i}as, Instituto de Ciencias Nucleares, \\
Universidad Nacional Aut\'onoma de M\'exico,
\\ Apartado Postal 70-543, CDMX 04510, M\'exico\\
 \vspace{0.2cm}
$^{\diamond}$
Institute for Theoretical Physics,\\
University of Amsterdam,\\
Science Park 904, 1098 XH Amsterdam, Netherlands\\
\vspace{0.2cm}
{\tt ricardo.espindola@correo.nucleares.unam.mx, \\
 alberto@nucleares.unam.mx, jpedraza@uva.nl}
\end{center}


\begin{center}
{\bf Abstract}
\end{center}
\noindent
In the holographic correspondence, subregion duality posits that knowledge of the mixed state of a finite spacelike region of the boundary theory allows full reconstruction of a specific region of the bulk, known as the entanglement wedge. This statement has been proven for local bulk operators. In this paper, specializing first for simplicity to a Rindler wedge of AdS$_3$, we find that generic \emph{curves} within the wedge are in fact \emph{not} fully reconstructible with entanglement entropies in the corresponding boundary region, even after using the most general variant of hole-ography, which was recently shown to suffice for reconstruction of arbitrary spacelike curves in the Poincar\'e patch. This limitation is an analog of the familiar phenomenon of entanglement shadows, which we call `entanglement shade'. We overcome it by showing that the information about the nonreconstructible curve segments is encoded in a slight generalization of the concept of entanglement of purification, whose holographic dual has been discussed very recently. We introduce the notion of `differential purification', and demonstrate that, in combination with differential entropy, it enables the complete reconstruction of all spacelike curves within
an arbitrary entanglement wedge in any 3-dimensional bulk geometry.
\vspace{0.2in}
\smallskip

\end{titlepage}

\tableofcontents

\section{Introduction and Conclusions}

In the quest to understand the holographic \cite{malda,gkp,w} emergence of a dynamical bulk spacetime out of degrees of freedom living on a lower-dimensional rigid geometry, much progress has originated from the Ryu-Takayanagi relation \cite{rt,hrt,lm,dlr}
\begin{equation}\label{rt}
S_A=\frac{\cA(\Gamma_A)}{4G_N}~.
\end{equation}
Here $S_A$ denotes the entanglement entropy of a spacelike region $A$ in the boundary theory: $S_A\equiv-\tr(\rho_{A}\ln\rho_{A})$, with $\rho_{A}\equiv\tr_{A^\mathsf{c}}\rho$
the reduced density matrix associated with $A$, or more precisely, with the domain of dependence of $A$ in the boundary theory, denoted $\cD_A$. $\cA(\Gamma_A)$ in (\ref{rt}) is the area\footnote{The connection with area applies when the bulk theory is classical Einstein gravity. For generalizations, see \cite{dong,camps,bdhm,flm,ew}.} of
the extremal codimension-two bulk surface $\Gamma_A$ that is homologous to $A$ (with $\p \Gamma_A =\p A$).

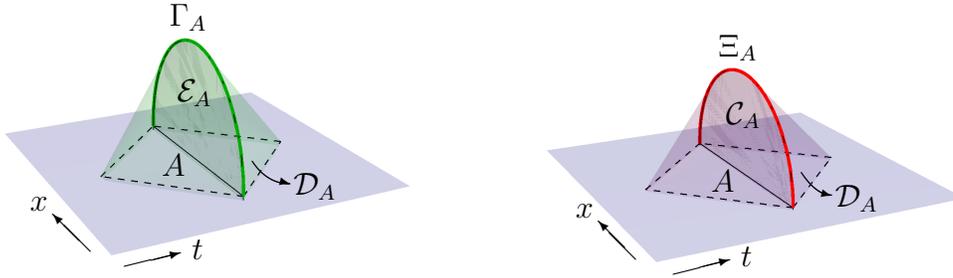
\begin{figure}[!ht]
\begin{center}
\vspace{-1cm}
\hspace*{-1cm}
  \includegraphics[width=8cm]{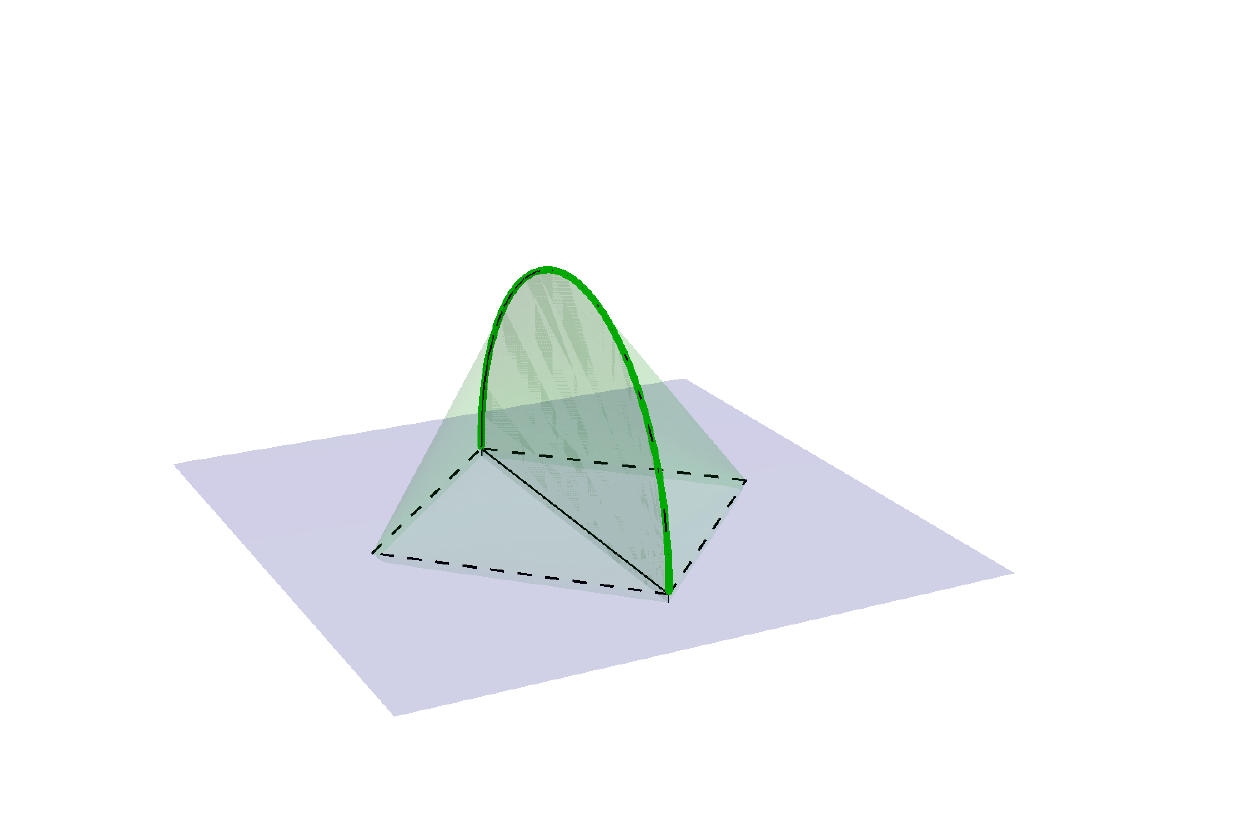}
  \hspace*{-1cm}
  \includegraphics[width=8cm]{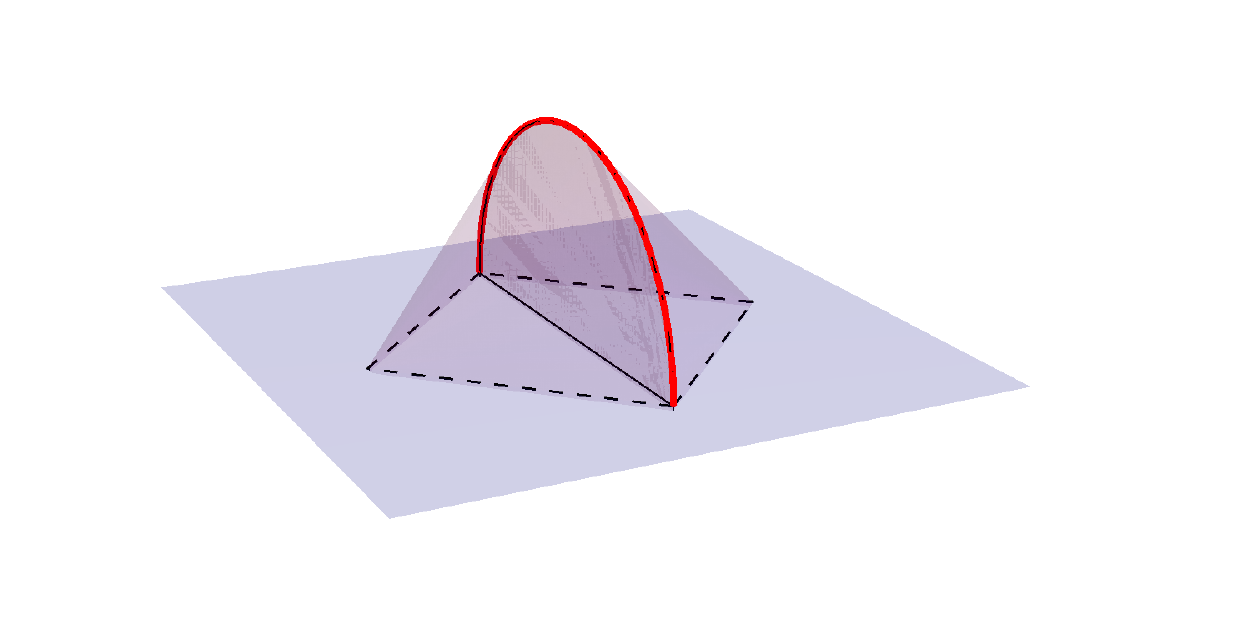}
  \hspace*{-1cm}
  \setlength{\unitlength}{1cm}
\begin{picture}(0,0)
\put(-4,2.5){$\cC_{A}$}
\put(-11.3,2.8){$\cE_{A}$}
\put(-4.1,3.4){$\Xi_{A}$}
\put(-11.4,3.8){$\Gamma_{A}$}
\put(-4.2,1.6){$A$}
\put(-11.5,1.8){$A$}
\qbezier(-3,1.85)(-2.8,1.55)(-2.6,1.55)
\put(-2.6,1.55){\vector(4,-1){0.01}}
\put(-2.55,1.4){$\cD_{A}$}
\qbezier(-10.3,2)(-10.1,1.7)(-9.8,1.7)
\put(-9.8,1.7){\vector(4,-1){0.01}}
\put(-9.75,1.55){$\cD_{A}$}
\put(-4.7,0.5){\vector(4,1){0.75}}
\put(-3.8,0.6){$t$}
\put(-5.2,0.7){\vector(-1,1){0.5}}
\put(-6,1.25){$x$}
\put(-12,0.6){\vector(4,1){0.75}}
\put(-11.1,0.7){$t$}
\put(-12.45,0.8){\vector(-1,1){0.5}}
\put(-13.25,1.35){$x$}
\end{picture}
\end{center}
\vspace*{-0.8cm}
\caption{Schematic depiction of the entanglement wedge $\cE_{A}$ and causal wedge $\cC_{A}$ for a boundary subregion $A$, or equivalently, for its boundary domain of dependence $\cD_A$. See the main text for the explicit definitions. For arbitrary bulk geometries, the entanglement wedge, bounded by null geodesics that are shot towards the boundary from the Ryu-Takayanagi surface $\Gamma_A$,
is larger than the causal wedge, bounded by null geodesics that are shot into the bulk from the edge of $\cD_A$. The spatial surface $\Xi_A$ on which the latter geodesics intersect is the causal information surface defined in \cite{hr}.
In a few situations $\Xi_A=\Gamma_A$, and the two types of wedges coincide. This happens in particular when $A$ is a spherical region in the vacuum of a $d$-dimensional conformal field theory, which for $d=2$ gives rise to the anti-de-Sitter-Rindler wedge considered throughout most of this paper.
\label{entanglementwedgefig}}
\end{figure}

Relation (\ref{rt}) informed in particular the idea of subregion duality \cite{vr,bousso1,hr,densitymatrix,bousso2,maulik}, and more specifically, the conjecture \cite{densitymatrix,wall,hhlr} that knowledge of $\rho_{A}$ allows full reconstruction of the entanglement wedge of $A$, denoted $\cE_{A}$ and defined as the domain of dependence of any codimension-one bulk spacelike region bounded by $\Gamma_A$ and $A$. See Fig.~\ref{entanglementwedgefig}a. An interesting property of the entanglement wedge is that it
is generally larger  \cite{wall,hhlr} than the bulk region that is causally accessible from $\cD_A$ (i.e., the intersection in the bulk of the causal past and the causal future of $\cD_A$), known as the causal wedge of $A$, and denoted $\cC_{A}$. See Fig.~\ref{entanglementwedgefig}b.

For a quantum field theory with a holographic dual, the large-$N$ and strong-coupling regime corresponds to the situation where the bulk theory is well approximated by Einstein gravity coupled to a small number of light local fields. Each of these fields $\phi$ (including the metric fluctuation $h_{mn}$) is dual to a simple local operator $\cO$ in the boundary theory. In this context, one aspect of reconstruction is being able to write the boundary counterpart of the bulk field operator $\phi$ placed at any given bulk point $x^m\equiv(x^{\mu},r)$, with $\mu$ running over the boundary directions, and $r$ the radial direction. This was first achieved with the well-known HKLL prescription \cite{bdhm2,bena,hkll,hmps,morrison}, which (at least for ball-shaped $A$) allows one to define $\phi(x,r)$ in $\cC_A$ by smearing $\cO(x)$ over $\cD_A$.
Using the connection with quantum error correction \cite{adh}, a proof was given in \cite{dhw} that local operators (acting within a code subspace) can in fact be reconstructed inside the full entanglement wedge $\cE_A$. See also \cite{jlms,fl,kim}, as well as the recent reviews \cite{dejonckheere,harlow}.

A different aspect of reconstruction is to be able to directly encode bulk curves or surfaces in terms of boundary data. This question was first addressed in \cite{hole-ography} for the case of global 3-dimensional anti-de Sitter spacetime (AdS$_3$), where an extremal surface $\Gamma_A$ is a geodesic, and its area $\cA(\Gamma_A)$ is a length. It was shown in that work that a generic bulk curve at fixed time, $x^m(\lambda)$ (with $\lambda$ an arbitrary parameter), can be represented by a family of intervals $I(\lambda)$ in the boundary theory, and a specific combination of the corresponding entanglement entropies $S_{I(\lambda)}$, known as the differential entropy $E$, yields the length of the curve.\footnote{A direct information-theoretic interpretation of $E$ within the boundary theory was provided in \cite{bartekinformation}, and an elegant reformulation of the dual bulk prescription was worked out in \cite{integralgeometry}, employing integral geometry.}  In this approach, known as hole-ography, the intervals $I(\lambda)$ are identified by the fact that their associated bulk geodesics $\Gamma_{I(\lambda)}$ are tangent to the bulk curve. By shrinking the curves to zero size, one can obtain in particular the most basic ingredients of the bulk geometry, points and distances, in terms of the pattern of entanglement in the given state of the field theory \cite{nutsandbolts}. Hole-ography thus provides direct access to the spacetime on which local bulk operators are to be placed, and therefore conceptually underlies the approach summarized in
 the previous paragraph. This is consistent with the fact that, purely within the field theory, entanglement is the more fundamental substrate from which correlators of local operators arise  \cite{hastings}. Hole-ography was examined in bulk dimensions higher than three in \cite{myers,cds,hmw}, and generalized to the case of time-dependent spacelike curves in \cite{hmw}. Other extensions can be found in \cite{veronikaresidual,taylorflavors,keeler,poincarepoint,bartekmodular}.

\begin{figure}[t]
\begin{center}
  \includegraphics[width=4cm]{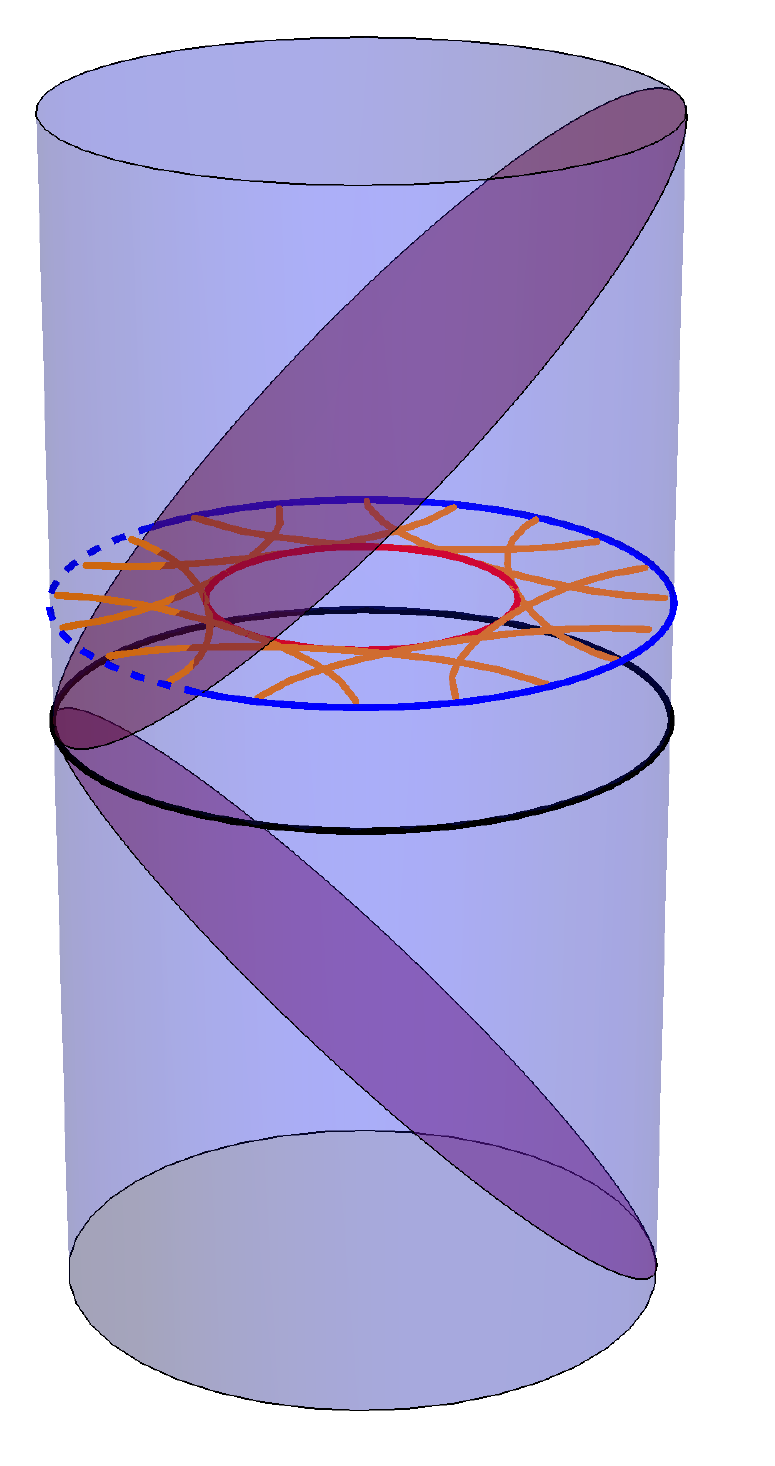}
  \hspace*{2cm}
  \includegraphics[width=4cm]{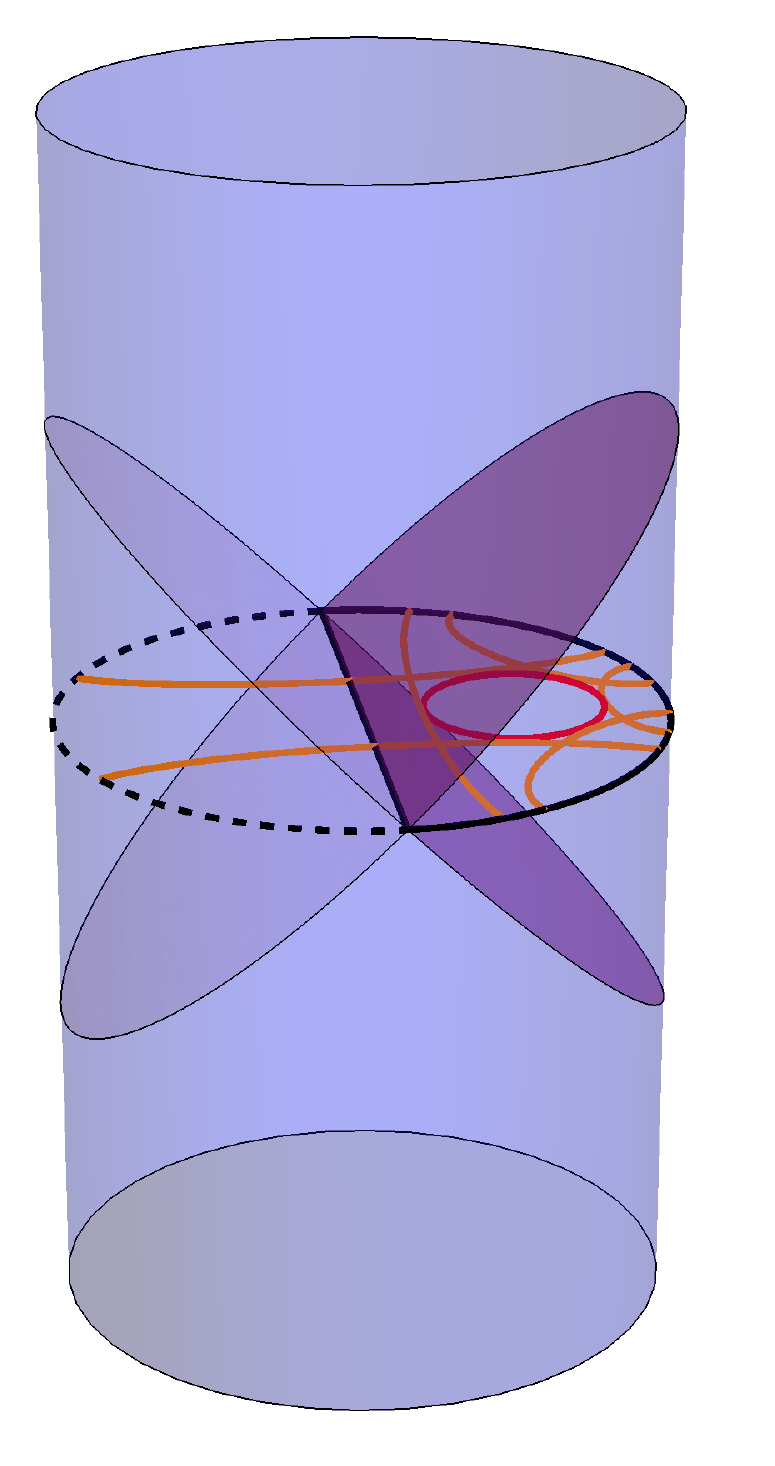}
  \setlength{\unitlength}{1cm}
\begin{picture}(0,0)
\put(-7.6,3.2){\vector(2,1){0.1}}
\put(-7.4,3.1){$x$}
\qbezier(-8.0,3.1)(-7.8,3.1)(-7.6,3.2)
\put(-7.4,5.2){\vector(3,1){0.1}}
\put(-7.3,5.1){$r$}
\qbezier(-7.4,5.2)(-7.9,5.1)(-8.3,4.9)
\put(-7.8,5.3){\vector(1,2){0.1}}
\put(-7.6,5.5){$t$}
\qbezier(-7.7,2)(-8.8,3.7)(-7.7,5.5)
\put(-10.7,6.0){\vector(0,1){0.5}}
\put(-10.8,6.7){$\tau$}
\put(-11.4,3.7){$\tau=0$}
\put(-9.8,6.4){\vector(1,0){0.1}}
\put(-9.6,6.2){$\theta$}
\qbezier(-10.3,6.6)(-10.0,6.4)(-9.8,6.4)
\put(-8.6,7.0){\vector(2,1){0.4}}
\put(-8.1,7.1){$\varrho$}
\put(-8.6,7.0){\circle*{0.1}}
\put(-1.3,3.2){\vector(2,1){0.1}}
\put(-1.5,2.8){$\bx$}
\qbezier(-1.7,3.1)(-1.5,3.1)(-1.3,3.2)
\put(-1.0,4.3){\vector(3,1){0.1}}
\put(-1.8,4.1){$\br$}
\qbezier(-1.0,4.3)(-1.6,4.1)(-2.0,3.9)
\put(-1.04,4.43){\vector(2,3){0.1}}
\put(-1.3,4.55){$\bt$}
\qbezier(-1.0,3.0)(-1.6,3.7)(-1.0,4.5)
\put(-4.4,6.0){\vector(0,1){0.5}}
\put(-4.5,6.7){$\tau$}
\put(-3.5,6.4){\vector(1,0){0.1}}
\put(-3.3,6.2){$\theta$}
\qbezier(-4.0,6.6)(-3.7,6.4)(-3.5,6.4)
\put(-5.1,3.7){$\tau=0$}
\put(-2.3,7.0){\vector(2,1){0.4}}
\put(-1.8,7.1){$\varrho$}
\put(-2.3,7.0){\circle*{0.1}}
\end{picture}
\end{center}
\vspace*{-0.8cm}
\caption{Each of these solid cylinders is a Penrose diagram for AdS$_3$, covered in full by the global coordinates $(\varrho,\tau,\theta)$, but only in part by the Poincar\'e coordinates $(t,x,r)$ on the left, or the Rindler coordinates $(\bt,\bx,\br)$ on the right. a) Generic spatial bulk curves in the Poincar\'e wedge (such as the circle shown in red) have segments whose tangent geodesics (shown in orange) are \emph{not} fully contained within the wedge. In spite of this, a variant of hole-ography that employs `null alignment' allows their reconstruction with entanglement entropies in the CFT \cite{poincarepoint}. b) A Rindler wedge covers a smaller portion of global AdS, and in particular, it does \emph{not} contain a full Cauchy slice. A priori, it is not clear if the `null alignment' variant of hole-ography is sufficient to reconstruct arbitrary bulk curves within the Rindler wedge (such as the circle shown in red).
\label{3dfig}}
\end{figure}
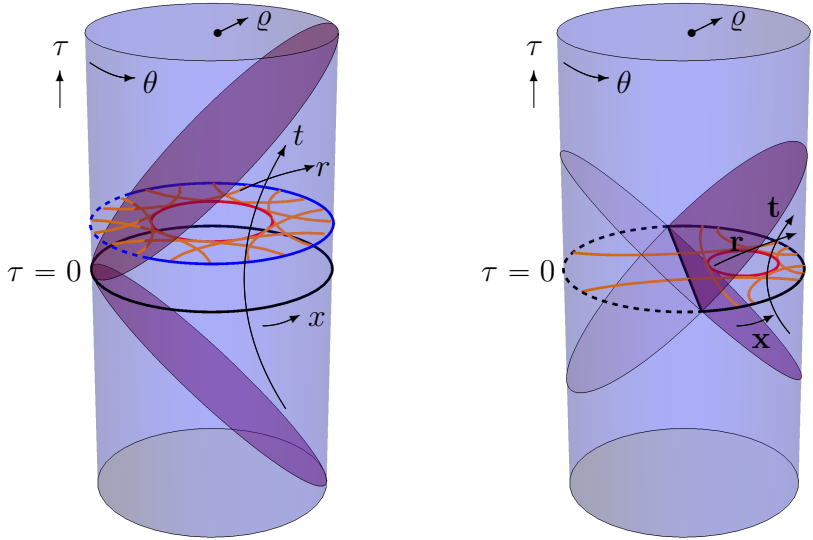

The simplest example of an entanglement wedge is the Poincar\'e patch of pure AdS$_3$, where $A$ is obtained by deleting a single point from the boundary circle that is at play in global AdS. In the most familiar presentation of Poincar\'e AdS/CFT, a conformal transformation is used to map this open interval to the whole real line, and the dual conformal field theory (CFT) then lives on 2-dimensional Minkowski spacetime. It was recently pointed out in \cite{poincarepoint} that in this setting hole-ography faces a serious challenge: generic bulk curves in the Poincar\'e wedge have segments whose tangent geodesics are \emph{not} fully contained within the wedge, meaning that they cannot be associated with entanglement entropies in the CFT. See Fig.~\ref{3dfig}a. This challenge was overcome in \cite{poincarepoint} by using a variant of hole-ography that employs `null alignment'. The key point, discovered in \cite{hmw}, is that $E$ reproduces the length of the curve even if the intervals $I(\lambda)$ are obtained not by shooting geodesics along the direction tangent to the curve, but along a new direction that has been shifted by a null vector orthogonal to the curve. It was shown in \cite{poincarepoint} that, for the segments of bulk curves that cannot be reconstructed with the standard prescription, it is always possible to reorient the geodesics in this manner to make sure they are contained within the Poincar\'e wedge, and therefore encode entanglement entropies. The conclusion then is that all spacelike curves in Poincar\'e AdS \emph{are} fully reconstructible.

Since the Poincar\'e patch has the special property that it contains a full Cauchy slice of global AdS, a question naturally arises: when considering a smaller entanglement wedge in AdS$_3$, e.g., an AdS-Rindler wedge, will null alignment again suffice to ensure reconstructibility of all spacelike bulk curves? This is the question that provides the initial motivation for this paper. The fact that AdS-Rindler is smaller than the Poincar\'e patch implies that there are fewer curves that need to be reconstructed, but on the other hand, there are more geodesics that exit the wedge. See Fig.~\ref{3dfig}b.

Our notation is established by writing the metric in the form (\ref{metricBH}). (The transformations from global AdS$_3$ to Poincar\'e and Rindler coordinates are given in Appendix~\ref{appendix}.) We begin by working out the explicit form of the AdS-Rindler geodesics in Section \ref{rindlersec}, first at constant time in \S~\ref{rindlerstaticsubsec} and then incorporating time dependence in \S~\ref{rindlergeneralsubsec}. With this information in hand, we identify in Section \ref{criterionstaticsubsec} a criterion for points on a static curve to be reconstructible using the standard tangent alignment. In Section \ref{criteriongeneralsubsec} we generalize this to time-dependent spacelike curves, incorporating the use of null alignment, defined in Eq.~(\ref{nullalignment}). The analysis reveals that a curve is reconstructible only if the two conditions (\ref{ineqUvUtau}) and (\ref{ineqsmallUrho}) are obeyed.
We then show in Section \ref{shadesubsec} that, \emph{even with null alignment, curves in an AdS-Rindler wedge generically have segments that cannot be reconstructed using entanglement entropies in the CFT}. The problem is that geodesics anchored on the boundary fail to reach certain bulk regions with a certain range of slopes. This obstruction is a generalization of the well-known phenomenon of entanglement shadows \cite{veronikaplateaux,ewshadows,entwinement,freivogel,ef,balasubramanian}, which we call \emph{entanglement shade}, and depict in Fig.~\ref{shadefig}.

Section \ref{mappingsubsec} delineates the problem in more detail, addressing the first step for reconstruction, which is to associate our bulk curve with a family of intervals in the boundary theory. We find that, while this can be done without difficulty for open curves that are not too steep (including those that can be shrunk down to describe points, as in \cite{nutsandbolts,poincarepoint}), all closed curves and generic open curves have some number of segments inside the entanglement shade, which by definition cannot be encoded in terms of boundary-anchored geodesics, and intervals in the CFT.

The resolution to the problem is developed in Section \ref{eopsubsec}. Focusing first on static curves, we show that the missing geodesics are closely related to the ones that have been very recently conjectured to describe the concept of entanglement of purification \cite{eop}, defined in \cite{takayanagi,phuc} and further explored in \cite{bao,takayanagi2,hty}. The relevant expressions can be seen in (\ref{eopdef}) and (\ref{eoph}). A slight generalization of this concept, given in (\ref{eopdef2}) and (\ref{eoph2}), requires one to find the optimal purification of the given mixed state, but then consider suboptimal bipartitions of the auxiliary degrees of freedom associated with that purification. \emph{We show that this variant of entanglement of purification enables the reconstruction of the problematic segments for static curves in an AdS-Rindler wedge. We then demonstrate that, with the help of null alignment, the prescription can be extended to time-dependent curves in the same wedge, and in fact,
to all spacelike curves within an arbitrary entanglement wedge $\cE_A$ in any 3-dimensional bulk geometry.}
This conclusion is our main result.
The task of reconstruction is completed explicitly in Section~\ref{esubsec}, where we show that, just like entanglement entropies can be combined to define the differential entropy (\ref{diffEs}) that yields the length of any curve segment outside the entanglement shade, entanglements of purification can be combined to define the \emph{differential purification} (\ref{diffP}) that reproduces the length of any segment inside the shade.

{}From the conceptual perspective, the crucial insight that emerges from \cite{takayanagi,phuc} and is reinforced by our results is that the optimization procedure involved in the calculation of the entanglement of purification identifies a specific set of additional field theory degrees of freedom $A'$, which in the gravity description live on the Ryu-Takayanagi surface $\Gamma_A$. After their addition, $\cE_A$ \emph{by itself} becomes dual to a pure state, and any curve segment can be encoded in terms of what is ultimately entanglement entropy in the enlarged version of the boundary theory.
As explained in Section \ref{eopsubsec},  at present the field theory interpretation of our recipe is completely clear only for the case where $A$ is connected,
or in the case where $A$ is disconnected but we restrict to bulk geometries with a moment of time-reflection symmetry, and curves located therein.
The remaining cases require a deeper understanding of the purified and excised version of subregion duality alluded to above. Even for the best understood cases, we would like to have better control over the explicit mapping between $A'$ and $\Gamma_A$ (for which both the `bit thread' picture of \cite{bitthread,bitthread2} and the results of the recent work \cite{wen} will probably be helpful), and the sense in which one should assign boundary conditions on $\Gamma_A$ for the bulk fields inside $\cE_A$. More generally, we need to understand in more detail the way in which bulk modular flow \cite{js,fl,jlms} implements time evolution for the purifying degrees of freedom $A'$ (and here again \cite{bitthread,bitthread2,wen} will likely be relevant).  Other important questions that we leave for future work are the generalization to bulk dimensions higher than 3 (which presumably should be possible at least under the symmetry conditions discussed for differential entropy in \cite{myers,cds,hmw}), and the connection between the hole-ographic method and other approaches to reconstruction
\cite{stereoscopy,guica,diamondography,blmms2,guica2,ksuw,kl,sw,fl,insideout,cp,verlinde,mnstw,no,no2,verlinde2,gt,knowlocation,eh1,eh2}.

\section{Rindler Geodesics}\label{rindlersec}

Starting with the vacuum of a CFT$_2$ on Minkowski space coordinatized by $(t,x)$, we take $A$ to be an interval of length $\ell$ at fixed time. Tracing over the degrees of freedom in the complement $A^\mathsf{c}$, the CFT state is described by the reduced density matrix $\rho_A$. The entanglement wedge of $A$, $\cE_A$, is an AdS-Rindler wedge, depicted in Fig.~\ref{3dfig}b in the special case where $A$ is at $t=0$ and runs from $x=-L$ to $x=L$, with $L$ the radius of curvature of AdS$_3$. To study $\cE_A$, it is convenient to work in a dimensionless coordinate system adapted to
the wedge (see Appendix \ref{appendix} for details), in which the metric takes the form
\be\label{metricBH}
ds^2=L^2\left(-\br^2d\bt^2+(1+\br^2)d\bx^2+\frac{d\br^2}{1+\br^2}\right)~.
\ee
Here $-\infty<\bt,\bx<\infty,0<\br<\infty$ cover $\cE_A$, and are directly analogous to the familiar Poincar\'e coordinates $(t,x,r)$: $\bt$ and $\bx$ run along the CFT directions, while
$\br$ labels the holographic direction.
In these coordinates, the minimal bulk surface $\Gamma_A$ has been mapped to the horizon at $\br=0$,
and the boundary is located at $\br\to\infty$. With the metric in the form (\ref{metricBH}), it is most natural to work directly with the dual CFT in the coordinates $(\bt,\bx)$, which can be related back to the original Minkowski coordinates through the conformal transformation (\ref{conformal}). In this description, the CFT state is thermal \cite{ch}.

\subsection{Geodesics at constant time} \label{rindlerstaticsubsec}

Without loss of generality, we can parametrize the geodesics in terms of $\bx$, using the two functions $\bt(\bx),\br(\bx)$.
Since the metric (\ref{metricBH})
is invariant under time translations, there is a class of geodesics at constant $\bt$. We will study these first. They are obtained by extremizing the proper length
\be
\mathcal{L}=\int d\bx\,\sqrt{1+\br^2+\frac{\br'^2}{1+\br^2}}~,
\ee
which leads to the equation
\be
(1+\br^2)\br''-\br\left[3\br'^2+(1+\br^2)^2\right]=0~.
\ee
The general solution, for a geodesic passing through the bulk point
$(\bx_{\mathrm b},\br_{\mathrm b}\equiv\br(\bx_{\mathrm b}))$ with slope $s\equiv \br'(\bx_{\mathrm b})$, is
\begin{equation}\label{staticgeodesic}
\br(\bx)=\frac{\br_{\mathrm b}(1+\br_{\mathrm b}^2)\cosh(\bx-\bx_{\mathrm b})+s\sinh(\bx-\bx_{\mathrm b})}{\sqrt{(1+\br_{\mathrm b}^2)^2
-s\br_{\mathrm b}(1+\br_{\mathrm b}^2)\sinh[2(\bx-\bx_{\mathrm b})]-(s^2+\br_{\mathrm b}^2(1+\br_{\mathrm b}^2)^2)\sinh^2(\bx-\bx_{\mathrm b})}}~.
\end{equation}
We can see from the numerator of this expression that static geodesics fall into two categories. If
\begin{equation}\label{nocross}
\br_{\mathrm b}^2(1+\br_{\mathrm b}^2)^2>s^2~,
\end{equation}
then $\br(\bx)$ vanishes nowhere, meaning that the geodesic does not reach the horizon. Both of its endpoints are then on the boundary, at the locations $\bx_{\pm}$ where the denominator in (\ref{staticgeodesic}) vanishes,
\begin{equation}\label{vpm}
\bx_{\pm}=\bx_{\mathrm b}+\frac{1}{2}\ln\left(\frac{2+s^2+5\br_{\mathrm b}^2+4\br_{\mathrm b}^4+\br_{\mathrm b}^6\pm2\sqrt{(1+\br_{\mathrm b}^2)^3(s^2+(1+\br_{\mathrm b}^2)^2)}}{
(s+\br_{\mathrm b}(1+\br_{\mathrm b}^2))^2}\right)~.
\end{equation}
The geodesic can be reexpressed in terms of these parameters as
\begin{equation}\label{staticgeodesicvpm}
\br(\bx)=\frac{\cosh\left(\bx-\frac{\bx_++\bx_-}{2}\right)}{
\sqrt{\sinh^2\left(\frac{\bx_+-\bx_-}{2}\right)-\sinh^2\left(\bx-\frac{\bx_++\bx_-}{2}\right)}}~.
\end{equation}
Alternatively, the geodesic can be written in terms of the location $(\bx_0,\br_0\equiv\br(\bx_0))$ of its point of closest approach to the horizon (where $\br'(\bx_0)=0$), given by
\be
\bx_0=\frac{\bx_++\bx_-}{2}\,,\qquad \br_0=\text{csch}\left(\frac{\bx_+-\bx_-}{2}\right)\,,
\ee
which can be inverted to obtain
\be\label{vpmstatic}
\bx_\pm=\bx_0\pm\text{arcsinh}\left(\frac{1}{\br_0}\right)\,.
\ee
The geodesic then takes the form
\be
\br(\bx)=\frac{\br_0\,\cosh(\bx-\bx_0)}{\sqrt{1-\br_0^2\sinh^2(\bx-\bx_0)}}\,.\label{rhostat}
\ee

The other category of static geodesics arises from considering bulk points $(\bx_{\mathrm b},\br_{\mathrm b})$ and slopes $s$ such that
\begin{equation}\label{cross}
\br_{\mathrm b}^2(1+\br_{\mathrm b}^2)^2<s^2~.
\end{equation}
In this case, the numerator of (\ref{staticgeodesic}) vanishes at
\begin{equation}\label{vh}
\bx_{\mathrm h}=\bx_{\mathrm b}-\arctanh\left(\frac{\br_{\mathrm b}(1+\br_{\mathrm b}^2)}{s}\right)~,
\end{equation}
meaning that the geodesic crosses the horizon at this location. Only one of the endpoints (\ref{vpm}) then lies on the boundary (the other one is outside the wedge, in the region $\br\to-\infty$). Denoting its location by $\bx_{\infty}$, the general expression (\ref{staticgeodesic}) can be presented in the form
\begin{equation}\label{staticgeodesicvhvinfty}
\br(\bx)=\frac{\csch(\bx_{\infty}-\bx_{\mathrm h})\sinh(\bx-\bx_{\mathrm h})}{\sqrt{1-\csch^2(\bx_{\infty}-\bx_{\mathrm h})\sinh^2(\bx-\bx_{\mathrm h})}}~,
\end{equation}
where it is evident that $\br$ vanishes at $\bx_{\mathrm h}$ and diverges at $\bx_{\infty}$.
It is easy to prove that there are no geodesics that cross the horizon \emph{twice}. This is in fact true in the static case for an entanglement wedge arising from a connected region $A$, on any background geometry, because it is guaranteed by the property of entanglement wedge nesting, i.e., $A\subset B\Rightarrow\cE_A\subset\cE_B$ \cite{densitymatrix,wall,nesting}.

Upon the requisite change of coordinates (see Appendix \ref{appendix}), one can check that the two categories of static Rindler geodesics obtained here agree with the planar limit of the global BTZ geodesics worked out in Section 6.1 of \cite{nutsandbolts}.

The cases with $\br_{\mathrm b}^2(1+\br_{\mathrm b}^2)^2=s^2$ lie precisely at the transition between the two categories (\ref{nocross}) and (\ref{cross}), so they belong to both, in the sense that they can be obtained as a smooth limit of geodesics in either category. For our purposes below, it is more convenient to assign them to the first category. When $\br_{\mathrm b}^2(1+\br_{\mathrm b}^2)^2=s^2$, we can see from (\ref{vpm}) that one of the endpoints of the geodesic lies at $\bx=\pm\infty$, so the length of a geodesic of this type encodes the entanglement entropy of a semi-infinite interval in the CFT. In the original CFT coordinates $(t,x)$ (related to $(\bt,\bx)$ through the conformal transformation (\ref{conformal})), this corresponds to an interval extending right up to the edge of the interval $A$ that gave rise to our Rindler wedge.

\subsection{Time-dependent geodesics contained within the Rindler wedge} \label{rindlergeneralsubsec}

The length of a time-dependent geodesic is given by
\be\label{lengthgeneric}
\mathcal{L}=\int d\bx\,\sqrt{1+\br^2(1-\bt'^2)+\frac{\br'^2}{1+\br^2}}\,.
\ee
We restrict our attention to spacelike geodesics, so
\be\label{bound}
\bt'^2<1+\frac{1}{\br^2}+\frac{\br'^2}{\br^2(1+\br^2)}~.
\ee
Extremizing (\ref{lengthgeneric}) we arrive, after some simplifications, at the following system of equations for $\br$ and $\bt$:
\begin{align}
(1+\br^2)\br''-\br\left[3\br'^2+(1+\br^2)^2(1-\bt'^2)\right]=0\,,\label{eomrho}\\
\br (1+\br^2) \bt ''+2 \br ' \bt '=0\,.\qquad\qquad\,\,\,\,\label{eomtau}
\end{align}

Just as in the static case examined in the previous subsection, there will be two categories of geodesics: those that have both endpoints at the boundary of the Rindler wedge, and those that cross the horizon. The novelty is that now the latter category includes as well geodesics that reach the horizon at \emph{both} ends. As will become clear in the following sections, for our purposes we will only need the geodesics of the first category, which are the ones that have an interpretation in terms of entanglement entropy in the CFT. To our knowledge, these time-dependent geodesics have not been written down in closed form in the previous literature.

To find the geodesics we proceed as follows. First, we solve equation (\ref{eomtau}) for $\bt'$:
\be
\bt'(\bx)= \bt_p \left(1+\frac{1}{\br^2}\right)\,,\label{taupsol}
\ee
where $\bt_p$ is an integration constant, which can be interpreted as the value of $\bt'$ at $\br\to\infty$. Notice that $\bt_p$ can be positive or negative, but its absolute value is bounded by condition (\ref{bound}).
For a geodesic of the type that interests us, having both endpoints on the boundary, the strictest bound on $|\bt_p|$ comes from the deepest point of the geodesic, $(\bt_0,\bx_0,\br_0)$, where $\br'=0$, so we obtain
\be\label{boundtaup2}
\bt_p^2<\frac{\br_0^2}{1+\br_0^2}\,.
\ee

Next, we plug (\ref{taupsol}) into (\ref{eomrho}) to obtain an equation for $\br$:
\be
\br^3(1+\br^2)\br''-\br^4\left[3\br'^2+(1+\br^2)^2\right]+\bt_p^2 (1+\br^2)^4=0\,.
\ee
The general solution for $\br$ is
\be
\br(\bx)=\sqrt{\frac{\bt_p^2(1+\br_0^2)^2+(\br_0^2 (1-\bt_p)-\bt_p)(\br_0^2(1+\bt_p)+\bt_p) \cosh^2(\bx-\bx_0)}{\br_0^2-(\br_0^2 (1-\bt_p)-\bt_p)(\br_0^2(1+\bt_p)+\bt_p)\sinh ^2(\bx-\bx_0)}}\,.\label{rhogen}
\ee
Finally, we plug (\ref{rhogen}) into (\ref{taupsol}), and integrate to obtain
\be\label{taugen}
\bt(\bx)=\bt_0+\text{arctanh}\left(\frac{\bt_p(1+\br_0^2)\tanh (\bx-\bx_0)}{\br_0^2}\right)\,.
\ee
As a consistency check, we can see that if we set $\bt_p=0$ we recover the static solution, with $\bt(\bx)=\bt_0$ and $\br(\bx)$ given by (\ref{rhostat}).

By solving for the values of $\bx$ where the denominator in (\ref{rhogen}) vanishes, it is easy to relate the 4 integration constants $(\bt_0,\bx_0,\br_0)$ and $\bt_p$ to the endpoint locations $(\bt_-,\bx_-)$ and $(\bt_+,\bx_+)$:
\begin{align}
\bx_\pm&=\bx_0\pm\text{arcsinh}\left(\frac{\br_0}{\sqrt{\br_0^4-\bt_p^2(1+\br_0^2)^2}}\right)\,,\label{vpm2}\\
\bt_\pm&=\bt_0\pm\text{arctanh}\left(\frac{\bt_p(1+\br_0^2)^{1/2}}{\br_0 \sqrt{\br_0^2-\bt_p^2(1+\br_0^2)}}\right)\label{taupm}\,,
\end{align}
or, equivalently,
\begin{align}\label{inversemaprho0:tdep}
\bx_0=\frac{\bx_++\bx_-}{2}&\,,\qquad \br_0=\frac{1}{\sqrt{1+\cosh^2\left(\frac{\bx_+-\bx_-}{2}\right)\text{sech}^2\left(\frac{\bt_+-\bt_-}{2}\right)}}\,.\\
\bt_0=\frac{\bt_++\bt_-}{2}&\,,\qquad \bt_p=\frac{\cosh\left(\frac{\bt_+-\bt_-}{2}\right)\sinh\left(\frac{\bt_+-\bt_-}{2}\right)}{\cosh\left(\frac{\bx_+-\bx_-}{2}\right)\sinh\left(\frac{\bx_+-\bx_-}{2}\right)}\,.\label{inversemaptaup:tdep}
\end{align}

Notice that (\ref{taupm}) is real as long as the bound (\ref{boundtaup2}) is obeyed, but in order for (\ref{vpm2}) to be real, we must require that
\be\label{boundtaup2b}
\bt_p^2<\frac{\br_0^4}{(1+\br_0^2)^2}~,
\ee
which is stronger than (\ref{boundtaup2}). If $\br_0^4/(1+\br_0^2)^2<\bt_p^2<\br_0^2/(1+\br_0^2)$, the geodesic is spacelike but bends towards the horizon at $\br=0$, so it does not belong to the class of geodesics examined in this subsection. We can solve for the remaining geodesics simply by relinquishing the use of $(\bt_0,\bx_0,\br_0,\bt_p)$ as parameters, but we will not write the explicit expressions because we will have no need for them in this paper.

\section{Criteria for Reconstructibility of Curves}\label{criterionsec}

\subsection{Static case} \label{criterionstaticsubsec}

Following \cite{hmw}, we will parametrize a bulk curve by functions $x^m(\lambda)=(\bt(\lambda),\bx(\lambda),\br(\lambda))$, where $\lambda$ runs from some initial value $\lambda_i$ to some final value $\lambda_f$. A hole in spacetime is demarcated by a curve that is closed, and consequently satisfies periodic boundary conditions, $x^m(\lambda_i)=x^m(\lambda_f)$. We can also consider open curves, which satisfy no such condition. In this subsection we will start by focusing on the simplest case: a curve at constant $\bt(\lambda)$. Unlike what happens in Poincar\'e-AdS \cite{poincarepoint}, in a Rindler wedge this static case is already nontrivial, because any slice at constant $\bt$ includes geodesics that exit the wedge (see Fig.~\ref{3dfig}b). On general grounds, therefore, we expect that there will be segments of the curve that cannot be reconstructed using tangent geodesics.

Given a static curve, our initial goal is to obtain the family of geodesics anchored on the boundary that are tangent to it at every point. The tangent vector is given by $u=(0,\bx'(\lambda),\br'(\lambda))$ and is spacelike everywhere. Since the metric is invariant under translations in $\bt$, the tangent geodesics will also lie on the constant-$\bt$ slice. We can directly use the results of Section \ref{rindlerstaticsubsec}: at any given point $\lambda$ on the bulk curve, the tangent geodesic is obtained by plugging $\bx_{\mathrm b}=\bx(\lambda)$, $\br_{\mathrm b}=\br(\lambda)$ and $s=\br'(\lambda)/\bx'(\lambda)$ into (\ref{staticgeodesic}). We know that this geodesic will have both of its endpoints on the boundary only if condition (\ref{nocross}) is obeyed, i.e., if
\be\label{criterium1}
\br(1+\br^2)>\left|\frac{d\br}{d\bx}\right|\,.
\ee
{\emph{ This then is our criterion for reconstructibility of constant-$\bt$ curves.}

On segments where (\ref{criterium1}) is violated, there is still the question of whether reconstruction can be achieved using null alignment \cite{hmw}. This means that, instead of shooting the desired geodesic along the tangent direction $u$, we shoot it along a new direction $U$ that has been shifted by a null vector orthogonal to the curve:
\begin{equation}\label{nullalignment}
U=u+n~,\qquad n\cdot n=0,\quad n\cdot u=0~.
\end{equation}
By construction, $U$ has the same norm as $u$, and the crucial fact is that, for any smooth choice of the function $n(\lambda)$, the differential entropy $E$  obtained with $U(\lambda)$ is the same as that obtained with $u(\lambda)$ \cite{hmw}. (For open curves, this requires addition of an appropriate $n$-dependent boundary term \cite{poincarepoint}.)

We want to know whether the possibility of reorienting geodesics as in (\ref{nullalignment}) is enough to guarantee the reconstructibility of segments whose tangent geodesics exit the Rindler wedge. In \cite{poincarepoint} it was shown that this is always true in the Poincar\'e wedge, for an infinite number of choices of $n(\lambda)$. The two explicit examples given in that work translate here into $n^\bt=-u^\bt$ (implying $U^{\bt}=0$) and $n^\br=-u^\br$ (implying $U^\br=0$). The first choice does not help here, where we have $u^\bt=0$ on account of our curve being static. If $n^\bt=0$, then the condition $n\cdot n=0$ yields the trivial solution $n=0$, and we have no way to satisfy (\ref{criterium1}). The difference with Poincar\'e is that static curves there had no nonreconstructible segments. One can likewise check that $n^\br=-u^\br$ does not work here.

It remains to determine if some other choice of $n(\lambda)$ can help. But while we do that, we might as well consider the general case where the curve is at varying $\bt$, because the calculations are essentially the same: once we add to $u(\lambda)$ a non-vanishing $n(\lambda)$, the geodesics under scrutiny will not be static.

\subsection{General case} \label{criteriongeneralsubsec}

Given an arbitrary spacelike curve $(\bt(\lambda),\bx(\lambda),\br(\lambda))$, we want to identify the geodesics that, instead of being tangent to it, are aimed along the vector $U(\lambda)$ defined in (\ref{nullalignment}). At each $\lambda$, we can specify the null vector $n(\lambda)$ by choosing a value for one of its components, say $n^\bt$, and then solving the two conditions $n\cdot n=n\cdot u=0$ for the remaining components. By doing so, we arrive at
\be\label{ngeneral}
  n=\left(n^\bt, n^\bt\frac{\br^2(1+\br^2)u^{\bt}u^\bx\pm \br u^{\br}|u|}{(u^{\br})^2+(1+\br^2)^2(u^{\bx})^2},
  n^\bt(1+\br^2)\frac{\br^2 u^{\bt}u^{\br}\mp\br(1+\br^2)u^{\bx}|u|}{(u^{\br})^2+(1+\br^2)^2(u^{\bx})^2} \right)~.
\ee
Notice that $n$ is determined by $n^{\bt}$ and a discrete choice of sign.

With $n$ in hand, we can construct the shifted vector $U\equiv u+n$, along which we wish to shoot our new geodesic. In Section \ref{rindlergeneralsubsec} we established that time-dependent geodesics which have both endpoints on the boundary of the Rindler wedge have the form (\ref{rhogen})-(\ref{taugen}). The four parameters $(\bt_0,\bx_0,\br_0)$ and $\bt_p$ determine the location of the endpoints $(\bt_-,\bx_-)$ and $(\bt_+,\bx_+)$ through the relations (\ref{vpm2})-(\ref{taupm}), which yield real values only if the bound (\ref{boundtaup2b}) is satisfied.

To ensure that our desired geodesic touches the bulk curve at the given point $(\bt(\lambda),\bx(\lambda),\br(\lambda))$, and has its tangent along $U(\lambda)$, we need to enforce the 4 conditions
\begin{gather}
\br=\sqrt{\frac{\bt_p^2(1+\br_0^2)^2+(\br_0^2 (1-\bt_p)-\bt_p)(\br_0^2(1+\bt_p)+\bt_p) \cosh^2(\bx-\bx_0)}{\br_0^2-(\br_0^2 (1-\bt_p)-\bt_p)(\br_0^2(1+\bt_p)+\bt_p)\sinh ^2(\bx-\bx_0)}}\,,
\label{rholambdag}\\
\bt=\bt_0+\text{arctanh}\left(\frac{\bt_p(1+\br_0^2)\tanh (\bx-\bx_0)}{\br_0^2}\right)\,,
\label{taulambdag}\\
\frac{U^\br}{U^\bx}=\frac{\br_0^2(1+\br_0^2)(\br_0^2 (1-\bt_p)-\bt_p)(\br_0^2(1+\bt_p)+\bt_p)\sinh[2(\bx-\bx_0)]}{2\sqrt{(1+\br_0^2)^2\bt_p^2
+(\br_0^4-(1+\br_0^2)^2\bt_p^2)\cosh^2(\bx-\bx_0)}\big[\br_0^2
-(\br_0^4-(1+\br_0^2)^2\bt_p^2)\sinh^2(\bx-\bx_0)\big]^{3/2}}\,,
\label{drholambdag}\\
\frac{U^\bt}{U^\bx}=\frac{\br_0^2 (1+\br_0^2)\bt_p\,\text{sech}^2(\bx-\bx_0)}{\br_0^4-(1+\br_0^2)^2\bt_p^2\tanh^2(\bx-\bx_0)}\,.\label{dtaulambdag}
\end{gather}
Our task is then to solve these four equations to determine the parameters of the geodesic.

Equation (\ref{taulambdag}) can be easily decoupled, since it is the only one with explicit dependence on $\bt_0$. From this equation we get
\be\label{tau0}
\bt_0=\bt-\text{arctanh}\left(\frac{\bt_p(1+\br_0^2)\tanh (\bx-\bx_0)}{\br_0^2}\right)\,,
\ee
which can be used once $\bx_0$, $\br_0$ and $\bt_p$ are known.

In order to solve the remaining three equations, we proceed as follows. First we solve (\ref{dtaulambdag}) for $\bt_p$~,
\be\label{taupsol2}
\bt_p=\frac{\br_0^2\left(\pm\sqrt{1+(U^\bt/U^\bx)^2\sinh^2[2(\bx-\bx_0)]}-1\right)}{2(1+\br_0^2)(U^\bt/U^\bx)\sinh^2(\bx-\bx_0)}\,.
\ee
We keep the two signs for now. Next, we plug (\ref{taupsol2}) into (\ref{rholambdag}) and solve for $\br_0$,
\be\label{rho0sol}
\br_0=\sqrt{\frac{2\br^2(U^\bt/U^\bx)^2\sinh^2(\bx-\bx_0)}{(1+\br^2)\left(\pm\sqrt{1+(U^\bt/U^\bx)^2\sinh^2[2(\bx-\bx_0)]}-1\right)-2\br^2(U_\bt/U_\bx)^2\sinh^2(\bx-\bx_0)}}\,.
\ee
At this point we notice that the option with minus sign is problematic, because it would make the argument of the square root in (\ref{rho0sol}) negative. We discard this option and keep the solution with the plus sign, both in (\ref{taupsol2}) and (\ref{rho0sol}). Next, we plug (\ref{taupsol2}) and (\ref{rho0sol}) into (\ref{drholambdag}) and obtain
\be\label{Urho/Uv}
\frac{U^\br}{U^\bx}=\frac{\br(1+\br^2)
\left(\sqrt{1+(U^\bt/U^\bx)^2\sinh^2[2(\bx-\bx_0)]}
-1-2(U^\bt/U^\bx)^2\sinh^2(\bx-\bx_0)\right)}{\left(\sqrt{1+(U^\bt/U^\bx)^2\sinh^2[2(\bx-\bx_0)]}-1\right)\tanh(\bx-\bx_0)}\,.
\ee

Given a point $(\bt(\lambda),\bx(\lambda),\br(\lambda))$ on the bulk curve, and a vector $U(\lambda)$ along which we wish to shoot a geodesic from there, we can determine the geodesic parameters by solving (\ref{Urho/Uv}) for $\bx_0$, and then using the result subsequently in (\ref{rho0sol}), (\ref{taupsol2}) and (\ref{tau0}). Proceeding in this way, we find
\begin{eqnarray}
\bx_0&=&\bx-\frac{1}{2}\arcsinh\left(\frac{2\br(1+\br^2) U^\bx U^{\br}}{\Delta}\right)~,
\label{v0sol2}\\
\br_0&=&\frac{\sqrt{-(U^{\br})^2
+\br^2(1+\br^2)^2\left[(U^{\bx})^2+(U^{\bt})^2+\br^2\left[(U^{\bx})^2-(U^{\bt})^2\right]\right]+\Delta}}
{\sqrt{2}\sqrt{(U^{\br})^2
+(1+\br^2)((U^{\bx})^2+\br^2\left[(U^{\bx})^2-(U^{\bt})^2\right])}}~,
\label{rho0sol2}\\
\bt_p&=&\frac{\br^2 U^{\bt}}{(1+\br^2)U^\bx}~,
\label{taupsol3}\\
\bt_0&=&\bt-\arctanh\left(\frac{-(U^{\br})^2+\br^2(1+\br^2)^2\left[(U^{\bx})^2-(U^{\bt})^2\right]-\Delta}
{2\br(1+\br^2)U^{\bt}U^{\br}}\right)~,
\label{tau0sol2}
\end{eqnarray}
where
\begin{equation}
\Delta\equiv\sqrt{\left(\br^2(1+\br^2)^2(U^{\bx}-U^{\bt})^2-(U^{\br})^2\right)
\left(\br^2(1+\br^2)^2(U^{\bx}+U^{\bt})^2-(U^{\br})^2\right)}~.
\label{Delta}
\end{equation}

Condition (\ref{boundtaup2b}) together with (\ref{taupsol3}) implies that the geodesic will have both of its endpoints on the boundary of our Rindler wedge only if
\begin{equation}\label{ineqUvUtau}
(U^\bx)^2-(U^{\bt})^2>0~.
\end{equation}
Additionally, in order for $(\bt_0,\bx_0,\br_0)$ to be real, we must demand that both factors inside the square root in (\ref{Delta}) are positive,\footnote{We cannot use the option where both factors are negative, because in that case (\ref{v0sol2}) is found \emph{not} to be a solution of Eq.~(\ref{Urho/Uv}).}
\begin{equation}\label{ineqsmallUrho}
\br^2(1+\br^2)^2(U^{\bx}-U^{\bt})^2-(U^{\br})^2>0
\quad\mbox{and}\quad
\br^2(1+\br^2)^2(U^{\bx}+U^{\bt})^2-(U^{\br})^2>0~.
\end{equation}
The first of these conditions implies the second if $U^\bx$ and $U^{\bt}$ have the same sign, while the reverse is true if $U^\bx$ and $U^{\bt}$ have opposite signs. By adding the two inequalities in (\ref{ineqsmallUrho}) we obtain
\begin{equation}\label{ineqsmallUrho2}
\br^2(1+\br^2)^2\left[(U^{\bt})^2+(U^{\bx})^2\right]-(U^{\br})^2>0~.
\end{equation}
Again, this is respectively implied by the first or second condition in (\ref{ineqsmallUrho}) if $U^\bx$ and $U^{\bt}$ have equal or opposite signs.

\emph{The inequalities (\ref{ineqUvUtau}) and (\ref{ineqsmallUrho}) are our two criteria for reconstructibility: segments where either one of these conditions is violated yield geodesics that are not associated with entanglement entropies in the CFT.} Condition (\ref{ineqUvUtau}) is directly analogous to the criterion found for the Poincar\'e wedge in \cite{poincarepoint}: it states that the projection of $U$ to the boundary ought to be spacelike. (This is more stringent than the requirement that $U$ itself be spacelike,
$-\br^2(1+\br^2)(U^{\bt})^2+(1+\br^2)^2(U^\bx)^2+(U^{\br})^2>0$, which is implied by (\ref{ineqUvUtau}) regardless of the value of $\br$.) Condition (\ref{ineqsmallUrho}), on the other hand, had no analog in Poincar\'e.

If we set $n^{\bt}=0$, then $U=u$, so we are back in the standard case of tangent alignment, and the segments where either (\ref{ineqUvUtau}) or (\ref{ineqsmallUrho}) are violated are then the ones that cannot be reconstructed using the original recipe for hole-ography \cite{hole-ography}. Notice in particular that in the static case, where $u^{\bt}=0$, condition (\ref{ineqUvUtau}) is satisfied automatically, and (\ref{ineqsmallUrho}) correctly reduces to (\ref{criterium1}). The latter connection shows that the existence of criterion (\ref{ineqsmallUrho}) is related to the fact that Rindler geodesics at constant $\bt$ do not cover the entire boundary of AdS.

We can go beyond tangent alignment by considering $n^{\bt}\neq 0$. Having incorporated into the analysis of this subsection the variant of hole-ography developed in \cite{hmw}, we can state in full generality that \emph{a segment on a (possibly time-dependent) spacelike curve will be reconstructible using null alignment only if there is some choice of $n^{\bt}(\lambda)$ and some choice of sign in (\ref{ngeneral}) such that both (\ref{ineqUvUtau}) and (\ref{ineqsmallUrho}) are satisfied.}

\subsection{Entanglement shade} \label{shadesubsec}

Now that we have understood the criteria for reconstructibility, the next logical step is to consider situations where the bulk curve at a given point is non-reconstructible via tangent alignment (either by violating (\ref{ineqUvUtau}) or (\ref{ineqsmallUrho}), or both), and try to show that it is always possible to choose a value of $n^{\bt}$ in (\ref{ngeneral}) to shift $u\to U\equiv u+n$ achieving reconstructibility. But upon attempting this, one is doomed to failure. In the case of the Poincar\'e wedge, examined in \cite{poincarepoint}, only the spacelike-projection condition analogous to (\ref{ineqUvUtau}) had to be satisfied, but for Rindler reconstruction we have in addition the small-slope condition (\ref{ineqsmallUrho}). One must show that there exists an $n^{\bt}$ such that \emph{both} inequalities are satisfied simultaneously, and in general this turns out not to be possible.

The factor of $\br^2$ in the positive term of (\ref{ineqsmallUrho}) indicates that it will be harder to reconstruct curves located in the vicinity of the horizon. To look for trouble in this region, assume that we are given a specific tangent vector $u$, and then proceed to expand our two conditions in a power series in $\br$, leaving $n^{\bt}$ in (\ref{ngeneral}) arbitrary. {}From (\ref{ineqUvUtau}) we obtain
\begin{equation}\label{spacelikesmallr}
(u^{\bx})^2-(u^{\bt}+n^{\bt})^2+\frac{2\sigma n^{\bt}u^{\bx}u^{\br}}{\sqrt{(u^{\bx})^2+(u^{\br})^2}}\br+O(\br^2)>0~,
\end{equation}
where $\sigma=\pm 1$ refers to the choice of sign in (\ref{ngeneral}). For $\br$ small enough that the $O(\br)$ term can be neglected, it is clear that $n^{\bt}$ can always be chosen for this inequality to be satisfied.  On the other hand, from either version of (\ref{ineqsmallUrho}) we obtain
\begin{equation}\label{notsteepsmallr}
-(u^{\br})^2+\frac{2\sigma n^{\bt}u^{\bx}u^{\br}}{\sqrt{(u^{\bx})^2+(u^{\br})^2}}\br+O(\br^2)>0~.
\end{equation}
If $\br$ is small enough that the first term dominates, we see that the inequality is always violated, regardless of the value of $n^{\bt}$.

There is a potential loophole in the preceding argument, because even if $\br$ is arbitrarily small, we could take $n^{\bt}$ to be arbitrarily large, and then the $O(\br)$ term in (\ref{notsteepsmallr}) cannot be neglected. Specifically, choosing
$$|n^{\bt}|>|u^{\br}/u^{\bx}|\sqrt{(u^{\bx})^2+(u^{\br})^2}/2\br$$
(and taking $n^{\bt}$ to have the same sign as $\sigma u^{\bx}u^{\br}$), we would ensure that the small-slope condition (\ref{notsteepsmallr}) is obeyed. But then when we consider the spacelike-projection condition (\ref{ineqUvUtau}) without any approximation,
\begin{eqnarray}
\left[-1+\frac{\br^2\left((\br+\br^3)u^{\bx}u^{\bt}+\sigma u^{\br}\sqrt{(u^{\bx})^2+\frac{(u^{\br})^2}{1+\br^2}+\br^2(-(u^{\bt})^2+(u^{\bx})^2)}\right)}{\left((1+\br^2)^2(u^{\bx})^2+(u^{\br})^2\right)^2}
\right](n^{\bt})^2&{}&\nonumber\\
-2\left[u^{\bt}-\frac{\br u^{\bx}\left((\br+\br^3)u^{\bx}u^{\bt}+\sigma u^{\br}\sqrt{(u^{\bx})^2+\frac{(u^{\br})^2}{1+\br^2}+\br^2(-(u^{\bt})^2+(u^{\bx})^2)}\right)}{(1+\br^2)^2(u^{\bx})^2+(u^{\br})^2}
\right]n^{\bt}&{}&\nonumber\\
-(u^{\bt})^2+(u^{\bx})^2&>&0~,\nonumber
\end{eqnarray}
we see that it is violated, because the first term $-(n^{\bt})^2$ dominates.

We have just shown that, given any vector $u$ with $u^{\br}\neq 0$, at sufficiently small $\br$ no orthogonal null vector $n$ exists such that the geodesic aimed along $U=u+n$ has both of its endpoints on the boundary of the Rindler wedge. In a similar fashion, one can show at all radial depths that sufficiently steep geodesics are problematic. More specifically, given any position $\br$, one finds that for $u$ with sufficiently large $u^{\br}$ no $n$ exists such that the geodesic aimed along $U=u+n$ is boundary-anchored. Our conclusion then is that, \emph{even using null alignment, bulk curves passing through certain points $(\bt,\bx,\br)$ with certain tangents $u$ cannot be reconstructed with entanglement entropies in the CFT.}

The obstruction we have found here, which prevents us from finding extremal curves in the bulk with certain made-to-order specifications, is analogous to the well-known occurrence
of entanglement shadows \cite{veronikaplateaux,ewshadows,entwinement,freivogel,ef,balasubramanian}.
The difference is that an entanglement shadow refers to a bulk region where boundary-anchored geodesics cannot penetrate, whereas here we find that boundary-anchored geodesics of a certain steepness cannot penetrate beyond a certain radial depth. In other words, whereas a shadow is a well-delineated subset of spacetime, the obstruction we are dealing with is present in a subset of the spacetime tangent bundle. The fact that the boundary of this region is not well-demarcated in spacetime proper motivates us to refer to this phenomenon as `entanglement shade', in contrast with shadow.\footnote{It might be useful for some readers to remember that a shadow is the dark silhouette cast by an object that blocks a source of light, whereas shade is a region of darkness of indefinite shape. The latter concept is normally used only when the source of light is the Sun.} The entanglement shade for the Rindler wedge is depicted in Fig.~\ref{shadefig}.\footnote{A preliminary discussion of the existence of entanglement shades can be found already in Section~7.1 of \cite{freivogel}, where they were referred to as `partial shadows'.}

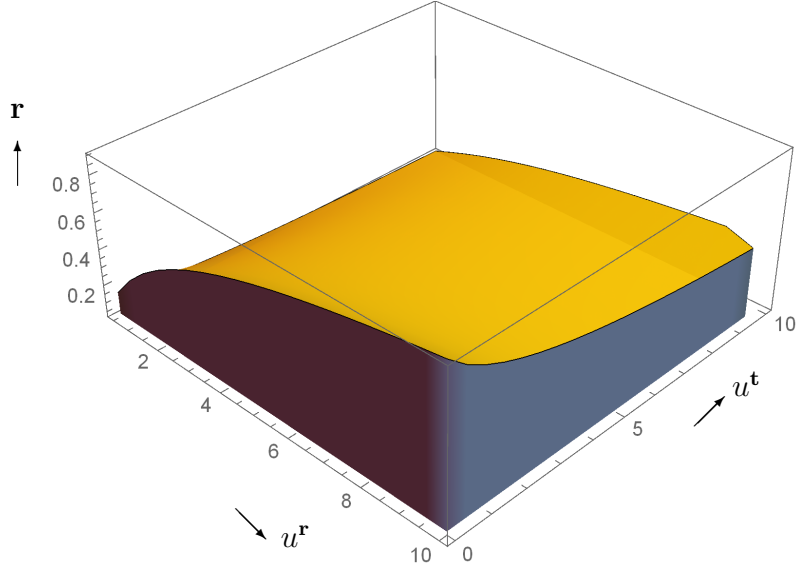
\begin{figure}[!hbt]
\begin{center}
  \includegraphics[width=10cm]{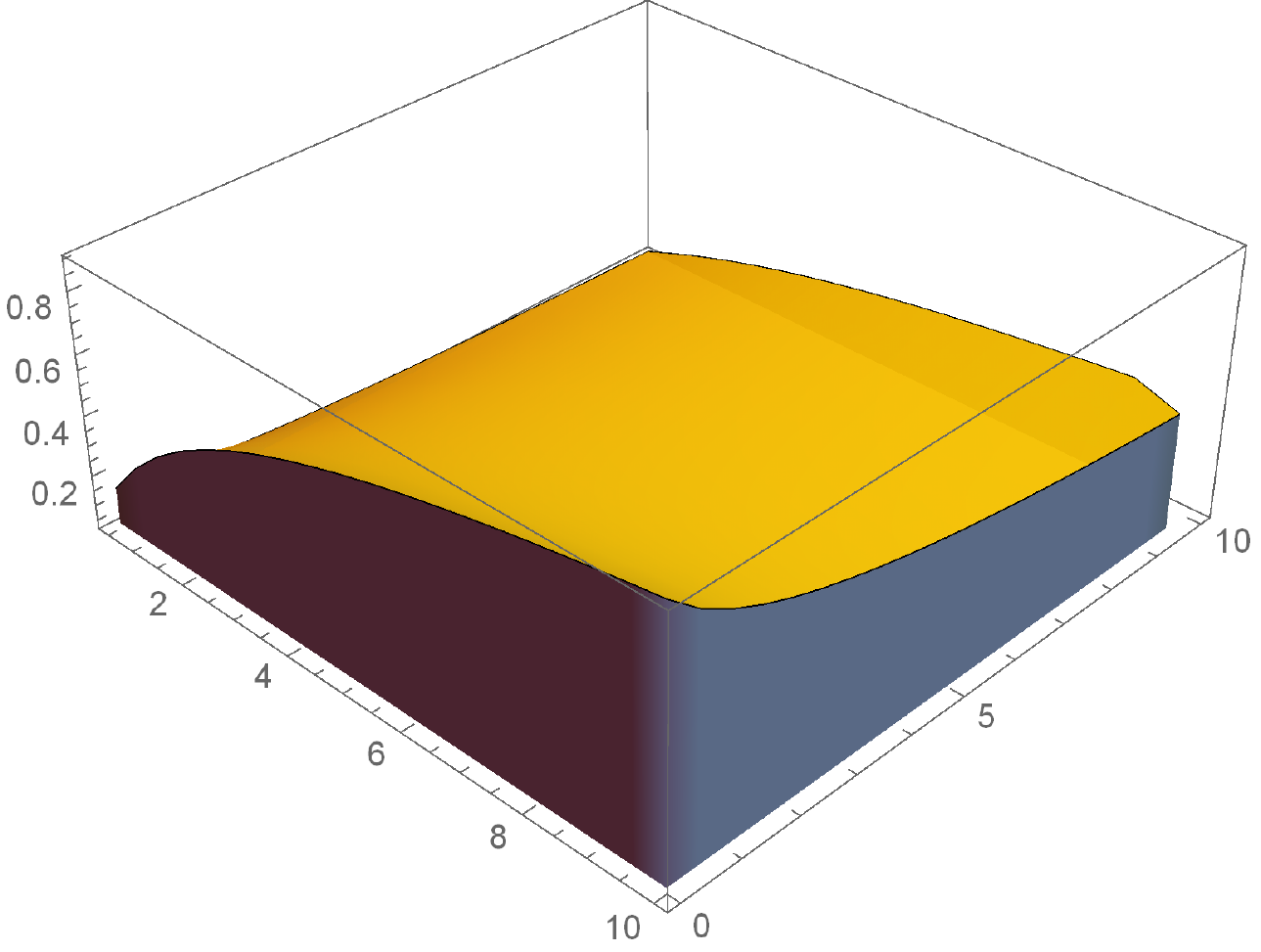}
  \setlength{\unitlength}{1cm}
\begin{picture}(0,0)
\put(-1.2,2.3){$u^{\bt}$}
\put(-1.7,1.9){\vector(1,1){0.4}}
\put(-7.2,0.3){$u^{\br}$}
\put(-7.8,0.8){\vector(1,-1){0.4}}
\put(-10.8,6.0){$\br$}
\put(-10.7,5.1){\vector(0,1){0.6}}
\end{picture}
\end{center}
\vspace*{-0.8cm}
\caption{Entanglement shade for a Rindler wedge in AdS$_3$, in the range $0<u^{\bt}<10$, $0<u^{\br}<10$, having chosen the parametrization $\lambda=\bx$ (which implies $u^{\bx}=1$). The shaded region indicates the radial depths that cannot be penetrated by geodesics with the indicated tangent vector $u$, or with any other vector $U$ obtained from it by null alignment ($U=u+n$ with $n\cdot n=n\cdot u=0$). As expected from the analysis in the main text, when we consider larger values of $u^{\br}$, corresponding to steeper curves, the shade grows larger. On the other hand, the figure shows that upon increasing the value of $u^{\bt}$ the shade is reduced. By symmetry, the radial position where the shade begins is independent of the sign of $u^{\bt}$ and $u^{\br}$, and of course, it is also independent of the values of $\bt$ and $\bx$. The entire region shown corresponds to spacelike $u$.
\label{shadefig}}
\end{figure}

\section{Full Reconstruction using Entanglement Entropy and Entanglement of Purification}\label{eopsec}

\subsection{Mapping bulk curves to CFT intervals} \label{mappingsubsec}

To reconstruct a spacelike bulk curve $\mathrm{C}$, the first step is to associate it with a family of intervals in the field theory. As stated before, we work with the CFT in the dimensionless coordinates $(\bt,\bx)$ appearing in the AdS-Rindler metric (\ref{metricBH}), which can be related back to the original Minkowski coordinates $(t,x)$ (where our Rindler wedge arose as the entanglement wedge for an interval of length $\ell$) through the conformal transformation (\ref{conformal}). For simplicity, we will focus for the most part on reconstruction of curves at constant $\bt$, described by the functions $\bx(\lambda)$, $\br(\lambda)$. As we will see, this case already contains the main novelty, and one additional trick will suffice to extend the prescription that we will develop to the case of $\bt$-dependent curves.

As in \cite{hmw,poincarepoint}, the family of CFT intervals $I(\lambda)$ that is associated with a given bulk curve is not unique: there is one family for each continuous choice of $n^{\bt}(\lambda)$ (and the sign) in (\ref{ngeneral}). The simplest possibility arises from the use of tangent alignment \cite{hole-ography}, which amounts to choosing $n^{\bt}(\lambda)=0$ for all $\lambda$. In this case we shoot geodesics $\Gamma_{I(\lambda)}$ along the vectors tangent to the curve, $u(\lambda)=(0,\bx'(\lambda),\br'(\lambda))$. On segments where $u(\lambda)$ is not too steep, in the sense that it obeys condition (\ref{criterium1}), the two endpoints of the geodesic lie on the boundary of the Rindler wedge, and therefore identify a specific interval $I(\lambda)$ in the CFT.  The remaining class of possibilities is to use null alignment \cite{hmw}, shooting the geodesics $\Gamma_{I(\lambda)}$ along $U(\lambda)\equiv u(\lambda)+n(\lambda)$ with $n(\lambda)$ given by (\ref{ngeneral}), for some choice $n^{\bt}(\lambda)\neq 0$. In this case, and also for time-dependent curves, the segments that are associated with intervals in the CFT are those where both the spacelike-projection condition (\ref{ineqUvUtau}) and the small-slope condition (\ref{ineqsmallUrho}) are satisfied. In both cases, the curve segments that fail to satisfy the relevant conditions are inside the entanglement shade described in Section \ref{shadesubsec}. For these segments there are no corresponding intervals in the CFT, and no encoding in terms of entanglement entropies.

If the curve $x^m(\lambda)$ is open and nowhere steep, it is completely outside of the shade, which implies that it is fully encoded by the family of intervals $I(\lambda)$. This includes both finite curves as in \cite{nutsandbolts,poincarepoint}, or infinite curves, in particular those that satisfy a periodicity condition at $\bx\to\pm\infty$, as in \cite{myers,hmw}. In the static case, the endpoints of the intervals are at the locations given by (\ref{vpm}),
\begin{equation}\label{Ilambda}
\bx_{\pm}(\lambda)=\bx+\frac{1}{2}\ln\left(\frac{2+\left(\frac{\br'}{\bx'}\right)^2
+5\br^2+4\br^4+\br^6\pm2\sqrt{(1+\br^2)^3\left[\left(\frac{\br'}{\bx'}\right)^2+(1+\br^2)^2\right]}}
{\left(\frac{\br'}{\bx'}+\br(1+\br^2)\right)^2}\right)~.
\end{equation}

The idea proposed in \cite{nutsandbolts}, of identifying any given bulk point as a `point-curve' obtained by shrinking a finite curve down to zero size, can be implemented in the Poincar\'e wedge by starting with an open curve whose slope at both endpoints is infinite, signaling that the curve becomes vertical there \cite{poincarepoint}. The resulting family $\Gamma_{I(\lambda)}$ comprises all geodesics that pass through the given point, and the centers of the corresponding intervals $I(\lambda)$ sweep the entire $x$-axis once. The analogous construction in the Rindler wedge involves an open curve whose slope at the endpoints, rather than being infinite, is on the verge of violating condition (\ref{criterium1}) (or (\ref{ineqsmallUrho})), meaning that the endpoints are at the edge of the entanglement shade. Upon shrinking such a curve down to zero size, we obtain all non-steep geodesics passing through the given bulk point $(\bt,\bx,\br)$. The corresponding intervals, if chosen to lie all at time $\bt$, are again those prescribed by (\ref{vpm}),
\begin{equation}\label{point}
\bx_{\pm}(s)=\bx+\frac{1}{2}\ln\left(\frac{2+s^2+5\br^2+4\br^4+\br^6\pm2\sqrt{(1+\br^2)^3(s^2+(1+\br^2)^2)}}{
(s+\br(1+\br^2))^2}\right)~.
\end{equation}
Here we are taking the point-curve to be parametrized by the slope $s$ of the geodesics involved in the final construction, which ranges from $-\br(1+\br^2)$ to $+\br(1+\br^2)$. These intervals sweep the entire $\bx$-axis once.

If the curve is closed, then it necessarily has at least two steep segments, where it enters the entanglement shade. The simplest possibility is exemplified by the circle in Fig.~\ref{circlefig}. For this type of closed curve, there are two non-steep segments $\mathrm{C}_1(\lambda)$, $\mathrm{C}_2(\lambda)$ that can again be associated, via (\ref{Ilambda}), with families of boundary-anchored geodesics $\Gamma_{I_1(\lambda)}$ and $\Gamma_{I_2(\lambda)}$ in the bulk, and with families of intervals $I_1(\lambda)$ and $I_2(\lambda)$ in the CFT. Each of these two families will cover the full $\bx$-axis once. These upper and lower segments of the curve are joined on the sides by two steep segments $\tilde{\mathrm{C}}_1(\lambda)$, $\tilde{\mathrm{C}}_2(\lambda)$. A generic closed curve will have $N\ge 2$ non-steep segments $\mathrm{C}_n(\lambda)$, alternating with $\tilde{N}=N$ segments $\tilde{\mathrm{C}}_n(\lambda)$ in the entanglement shade. The former will be associated with $N$ families of intervals $I_n(\lambda)$, each of which sweeps over the entire $\bx$-axis. The same is true for an open curve that has segments inside the shade, but in that case, the number $\tilde{N}$ of steep segments is not necessarily equal to $N$. For both closed and open curves, the situation is exactly analogous to the one described for the Poincar\'e wedge in \cite{poincarepoint}, with the difference being that in that setting the segments `inside' the entanglement shade are only those that are strictly vertical, which are normally isolated points for a generic curve.\footnote{We write `inside' in quotes because points or segments of curves in Poincar\'e that are vertical ($r'(\lambda)/x'(\lambda)\to\pm\infty$) are really at the \emph{edge} of the would-be shade. They can be described as limits of points or segments that are definitely outside the shade, and the corresponding geodesics encode the entanglement entropy of CFT intervals that are semi-infinite. For a Poincar\'e wedge, then, there is strictly speaking no (interior of the) entanglement shade \cite{poincarepoint}.}

\begin{figure}[!hbt]
\begin{center}
  \includegraphics[width=10cm]{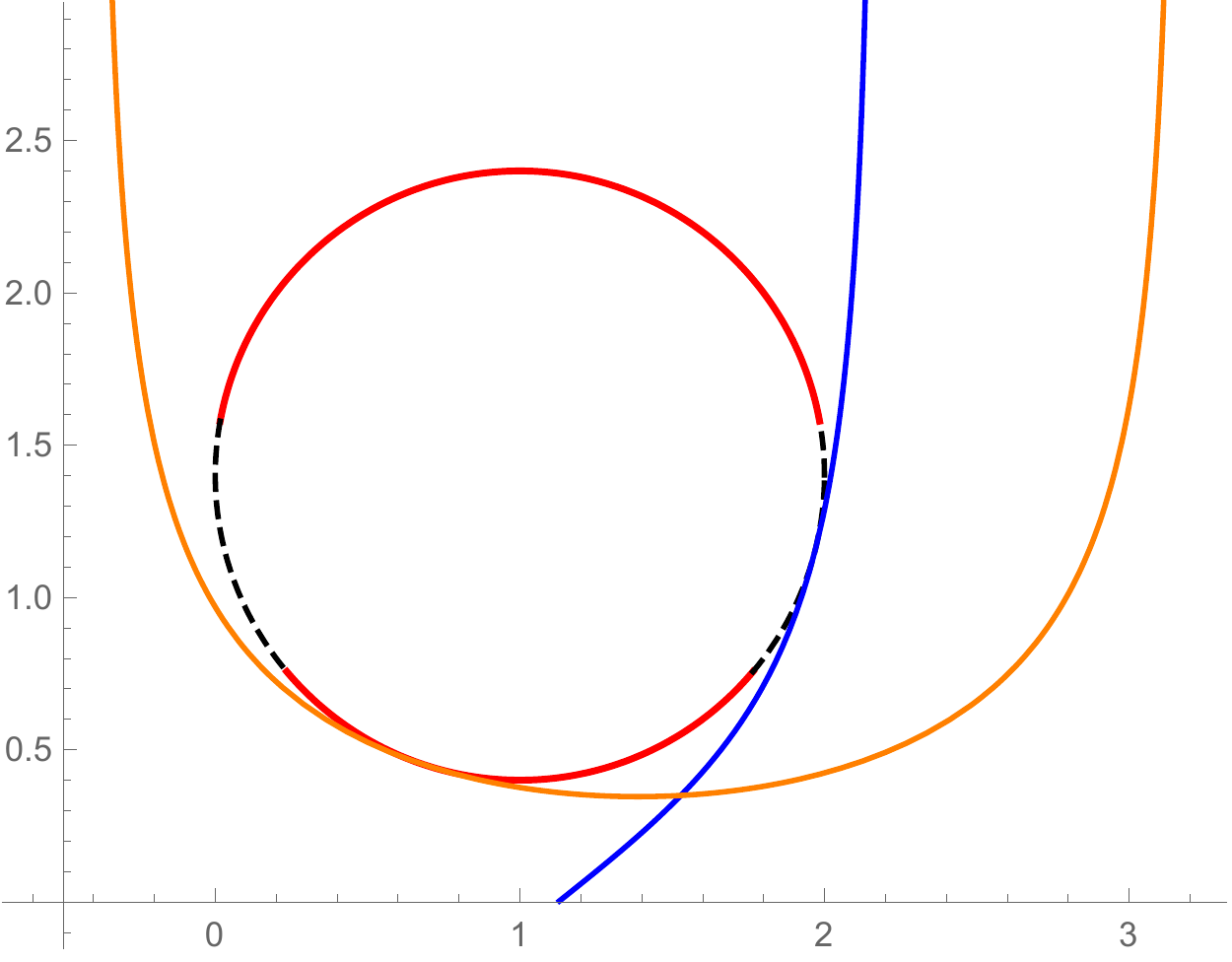}
  \setlength{\unitlength}{1cm}
\begin{picture}(0,0)
\put(-10.2,7.5){$\br$}
\put(0.0,0.2){$\bx$}
\put(-6.6,1.8){$\mathrm{C}_2(\lambda)$}
\put(-6.6,5.9){$\mathrm{C}_1(\lambda)$}
\put(-8.3,3.5){$\tilde{\mathrm{C}}_2(\lambda)$}
\put(-4.8,3.5){$\tilde{\mathrm{C}}_1(\lambda)$}
\end{picture}
\end{center}
\vspace*{-0.8cm}
\caption{An example of a closed spacelike curve: a circle at constant time $\bt$, centered at $\bx=1$, $\br=1.4$, with coordinate radius $a=1$. The top and bottom, shown in solid red, have tangent geodesics of the type (\ref{staticgeodesicvpm}), lying fully within the Rindler wedge. A sample such geodesic is shown in orange, with both of its endpoints extending up to the boundary at $\br\to\infty$. This is not true for the segments on the sides, shown in dashed black, which violate condition (\ref{criterium1}) and therefore cannot be reconstructed using entanglement entropies. Geodesics tangent to them, such as the one shown in blue, are of the type (\ref{staticgeodesicvhvinfty}), and have one endpoint on the boundary but cross the horizon $\br=0$ on the other side. If we parametrize the circle by $\lambda\in [0,1)$, with $\lambda=0$ located at the top, the gluing between the four segments occurs at the values $\lambda=0.138, 0.278, 0.722, 0.862$.
If we wished, we could use null alignment (\ref{nullalignment}) to reduce the size of the dashed segments, but as discussed in the main text, no choice of $n$ can make them disappear completely.
\label{circlefig}}
\end{figure}
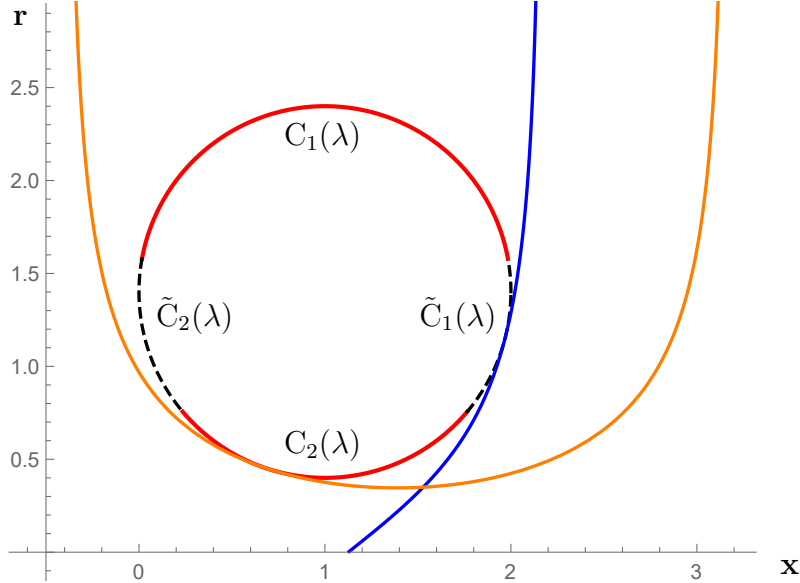

The upshot is that \emph{generic curves in the Rindler wedge $\cE_A$ contain some number $\tilde{N}$ of segments $\tilde{\mathrm{C}}_n(\lambda)$ in the entanglement shade, which cannot be encoded as intervals within the prescribed region $A$ of the CFT.} The corresponding geodesics (such as the blue geodesic in Fig.~\ref{circlefig}) have one endpoint on the boundary and one on the Rindler horizon, and are consequently not associated with entanglement entropies. As we will see in the next subsection, they are associated with a different measure of correlations, entanglement of purification \cite{eop}, whose holographic dual has been discussed in the very recent works \cite{takayanagi,phuc,bao,takayanagi2,hty}.

\subsection{Entanglement of purification} \label{eopsubsec}

The entanglement of purification is a measure of correlations, both quantum and classical, expressed in terms of entanglement of a certain pure state. In more detail, given a quantum system $A$ bipartitioned into sets of degrees of freedom $B$ and $C$ ($A=BC$), in a state described by a density matrix $\rho_{BC}$, we know that the von Neumann entropy $S_{BC}>0$ if the state is mixed. In that case, the entanglement entropies $S_{B}\neq S_{C}$ quantify quantum \emph{and} classical correlations between $B$ and $C$. A purification of this system is a set $A'$ of additional degrees of freedom, together with a choice of pure state $\ket{\psi}$ for the overall system $BCA'$, such that $\tr_{A'}\ket{\psi}\bra{\psi}=\rho_{BC}$. $S_{BC}$ is then understood as arising entirely from entanglement between $BC$ and $A'$. If we further partition the auxiliary system $A'$ into $B'$ and $C'$, we can compute instead the entanglement entropy $S_{BB'}=S_{CC'}$, which also arises purely from entanglement. By optimizing among all possible purifications and all possible partitions $B'C'$, the entanglement of purification between $B$ and $C$ is defined as \cite{eop}
\begin{equation}\label{eopdef}
\mathrm{P}(B:C)\equiv\min_{\scriptstyle \ket{\psi},B'} S_{BB'}~.
\end{equation}

In the holographic context, a plausible counterpart of $\mathrm{P}$
on the gravity side has been identified very recently \cite{takayanagi,phuc}. Given a time-independent bulk geometry dual to some state in a field theory, and a choice of spatial subsystem $A$ formed by two non-overlapping regions $B$ and $C$ at constant time $t$ on the boundary, we expect by subregion duality that the density matrix $\rho_{BC}$ encodes the portion of the bulk geometry inside the entanglement wedge $\cE_{BC}$  \cite{densitymatrix,wall,hhlr}. The slice of $\cE_{BC}$ at time $t$, which we will denote by  $\cE_{BC}|_t$, is bounded by $B$, $C$ and the minimal codimension-2 surface $\Gamma_{BC}$ (the Ryu-Takayanagi surface corresponding to $BC$). Within $\cE_{BC}|_t$~, we can find the minimal-area surface $\Sigma$ that ends on $\Gamma_{BC}$ and separates $B$ from $C$. The area of $\Sigma$ in Planck units,
\begin{equation}\label{eoph}
P(B:C)\equiv\frac{\cA(\Sigma)}{4G_N}~,
\end{equation}
was argued in \cite{takayanagi,phuc} to agree with the entanglement of purification (\ref{eopdef}),
$P=\mathrm{P}$.\footnote{We refrain from denoting the entanglement of purification by $E_p$ or $E_P$ as in \cite{eop,takayanagi,phuc,bao,takayanagi2,hty}, because $E$ is the symbol of choice for differential entropy \cite{hole-ography,bartekinformation,integralgeometry,nutsandbolts,myers,cds,hmw,veronikaresidual,taylorflavors,keeler,poincarepoint}, which we will be employing in the next subsection, and subscripts are used throughout this paper to refer to CFT intervals.} The construction is illustrated in Fig.~\ref{eopfig}a. In short, the holographic dual of the entanglement of purification is conjectured to be given by the minimal cross section of the entanglement wedge. This conjecture can be motivated by the tensor network interpretation of holography \cite{swingle,qi,happy,bartektensor,nothappy}, and the main evidence that supports it is the fact that
$P$
satisfies precisely the same inequalities as
$\mathrm{P}$. The extension to the case of overlapping $B$ and $C$ was put forward in \cite{bao}, and the generalization to the non-static setting was given in \cite{takayanagi,phuc}. The proposal has been explored further in \cite{bao,takayanagi2,hty}.

\begin{figure}[!htbp]
\begin{center}
  \includegraphics[width=4cm]{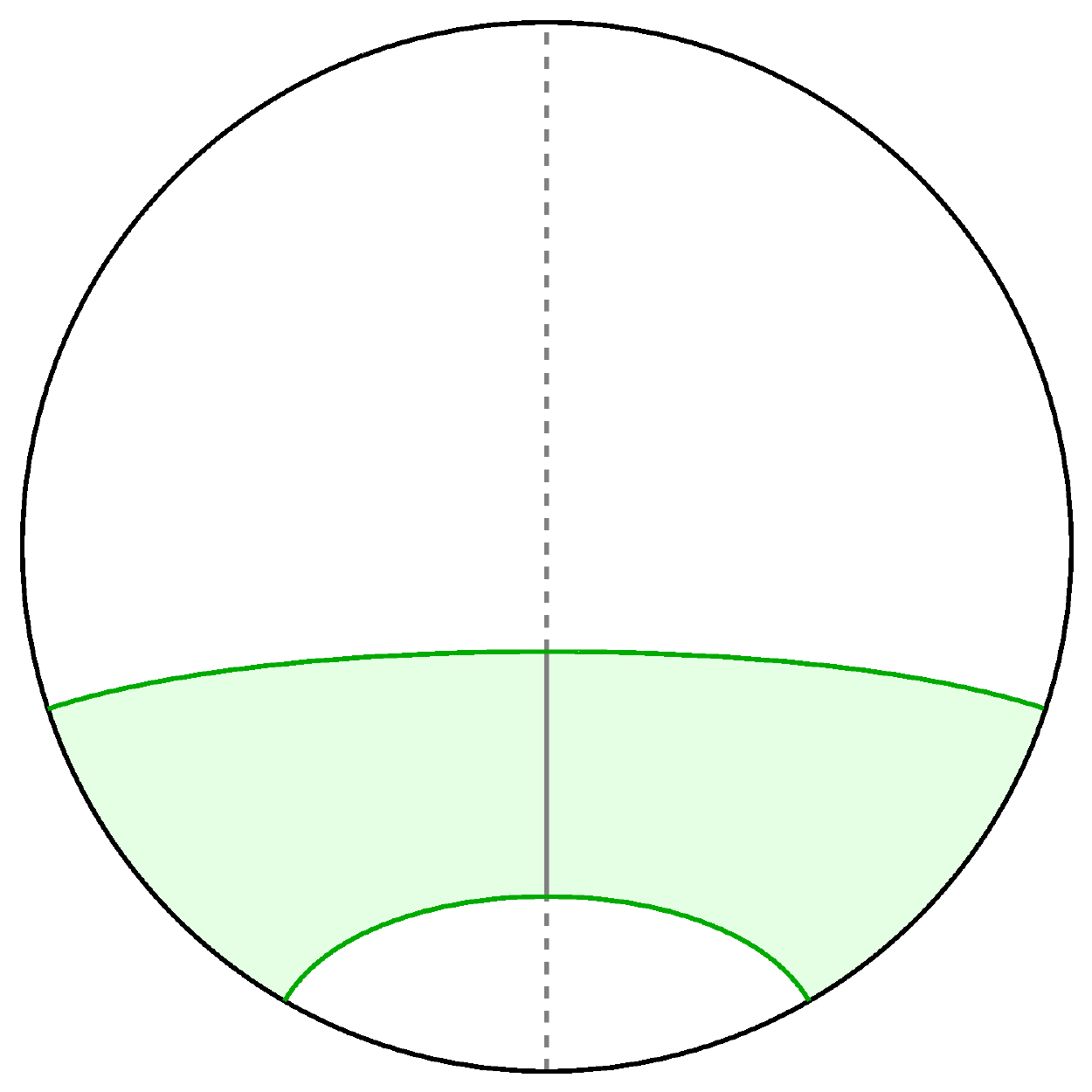}
  \hspace*{1cm}
  \includegraphics[width=4cm]{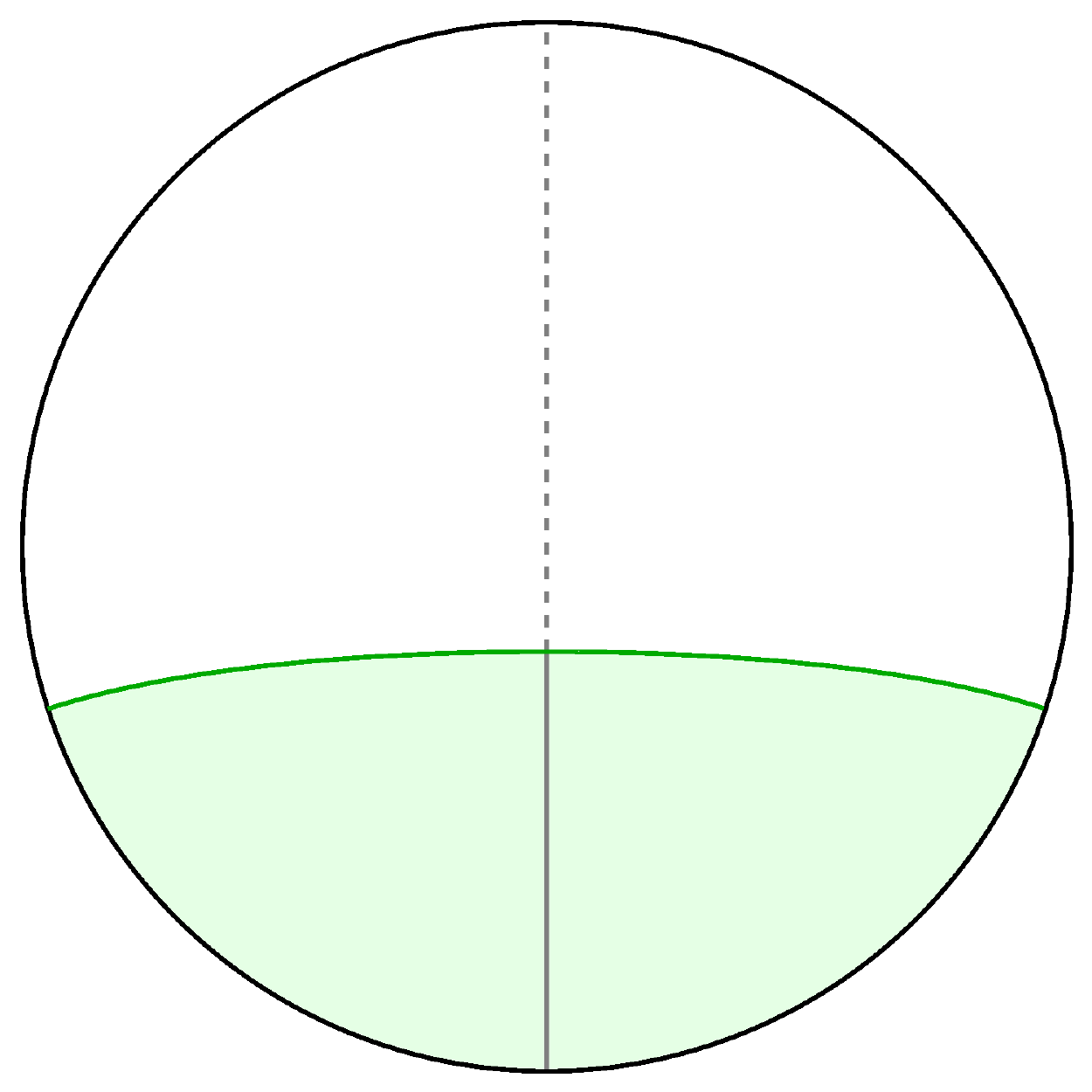}
  \hspace*{1cm}
  \includegraphics[width=4cm]{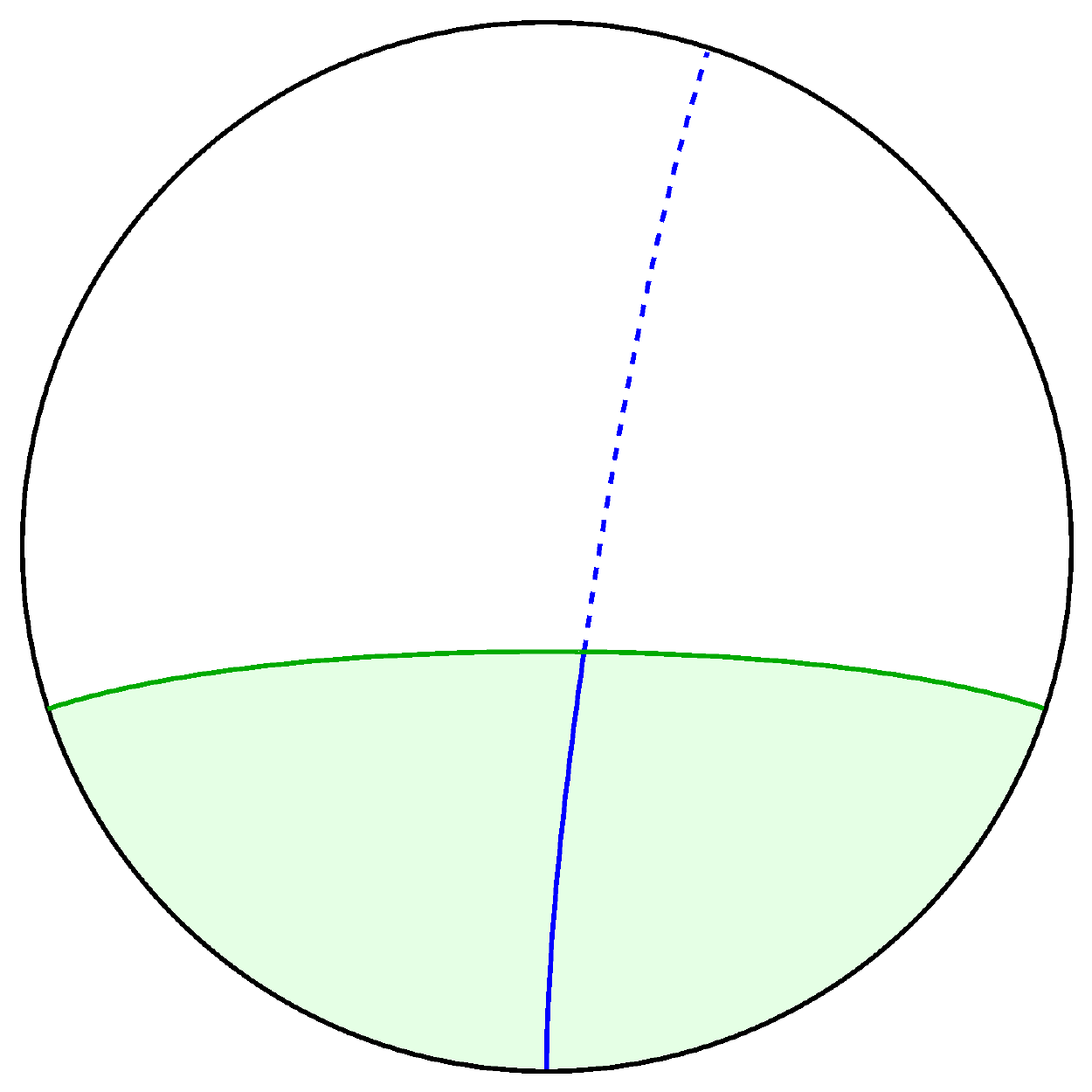}
  \setlength{\unitlength}{1cm}
\begin{picture}(0,0)
\put(-14.3,0.9){$\cE_{BC}|_t$}
\put(-9.0,0.9){$\cE_{BC}|_t$}
\put(-3.75,0.9){$\cE_{BC}|_t$}
\put(-12.7,1){$\Sigma$}
\put(-7.45,1){$\Sigma$}
\put(-2.1,1){$\Sigma'$}
\put(-14.85,0.6){$B$}
\put(-11.2,0.6){$C$}
\put(-13.4,0.3){$B'$}
\put(-12.65,0.3){$C'$}
\put(-13.9,1.7){$B'$}
\put(-12.15,1.7){$C'$}
\put(-9.35,0.4){$B$}
\put(-6.1,0.4){$C$}
\put(-8.6,1.7){$B'$}
\put(-6.85,1.7){$C'$}
\put(-4.1,0.4){$B$}
\put(-0.85,0.4){$C$}
\put(-3.35,1.7){$B'$}
\put(-1.6,1.7){$C'$}
\end{picture}
\end{center}
\caption{Ingredients for the holographic computation of the entanglement of purification $P$ and its generalization $P'$. The disk represents a constant-time slice of a static geometry dual to some pure state. Upon restricting the field theory to the region $A=BC$, we are left in the gravity description with the corresponding spatial slice of the entanglement wedge of $BC$, shown as the shaded region.  a) In the generic case where $B$ and $C$ are not contiguous, the Ryu-Takayanagi surface $\Gamma_{BC}$ has two disconnected components, indicated in green. Running between them at the narrowest part of the shaded region we see the entanglement wedge cross section, $\Sigma$, whose area encodes, according to (\ref{eoph}), the entanglement of purification (\ref{eopdef}) for the bipartition $BC$ of the given state. The corresponding minimal surface in the overall geometry would include the dotted segments as well, but these are excluded from the definition of $P$.
The degrees of freedom $A'$ of the purification `live on' $\Gamma_{BC}$, and $\Sigma$ partitions them into a specific choice of $B'$ and $C'$.
 b) In the particular case where $B$ and $C$ are contiguous, one of the components of $\Gamma_{BC}$ shrinks down to the transition point between $B$ and $C$, and $\Sigma$ is seen to extend from there to the closest point in the remaining, finite component. If the overall geometry is global AdS, the shaded region is an AdS-Rindler wedge.
 c) If in the setup of b) we consider instead a minimal surface $\Sigma'\neq\Sigma$, we obtain a different, suboptimal partition of $A'$ into $B'$ and $C'$, and the area of $\Sigma'$ is then expected to yield via (\ref{eoph2}) the entanglement of purification (\ref{eopdef2}) associated with that specific partition.
 \label{eopfig}}
\end{figure}
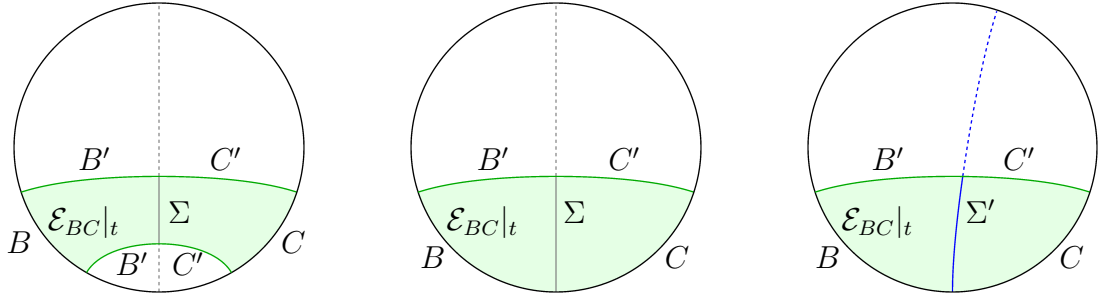

The connection with our story arises from considering the case where the bulk geometry is global AdS$_3$, and regions $B$ and $C$ are contiguous. As seen in Fig.~\ref{eopfig}b, the entanglement wedge for $A=BC$ is then our AdS-Rindler wedge $\cE_A$, and its minimal cross section $\Sigma$ is a geodesic that extends from the point on the boundary where $B$ and $C$ meet to the Rindler horizon $\Gamma_A$. This geodesic is of the type (\ref{staticgeodesicvhvinfty}), just like the blue geodesic in Fig.~\ref{circlefig}, and the other geodesics we were missing in the attempt in Section~\ref{mappingsubsec} of reconstructing curves using only entanglement entropies. But there is a difference between the two. $\Sigma$ in this context is determined exclusively by the location $\bx_{\infty}$ where $B$ and $C$ join, because it stretches from there to the \emph{closest} point on the Rindler horizon. This minimization condition uniquely determines the remaining parameter $\bx_{\mathrm{h}}$ in (\ref{staticgeodesicvhvinfty}). On the contrary, for geodesics tangent to curve segments $\tilde{\mathrm{C}}_n$ inside the entanglement shade, such as the blue geodesic in Fig.~\ref{circlefig}, $\bx_{\infty}$ and $\bx_{\mathrm{h}}$ are independent parameters, fixed by the two conditions that the geodesic passes through the given point on the bulk curve, $(\bx(\lambda),\br(\lambda))$, and that it has the required slope, $\br'(\lambda)/\bx'(\lambda)$.

What does this difference signify in the CFT language? To answer this question, let us first step back to notice from (\ref{eoph}) that
the optimal purification called for in (\ref{eopdef}) is \emph{not} the pure state dual to the entire bulk spatial slice in Figs.~\ref{eopfig}a or \ref{eopfig}b, which is what we had before restricting the CFT to region $A$. We know this because $\Sigma$ in (\ref{eoph}) does not run across the entire bulk, i.e., it does not include the dotted segments in Figs.~\ref{eopfig}a or \ref{eopfig}b. The exclusion of those segments indicates that the degrees of freedom $A'$ in the optimal purification are directly taken to `live on' the Ryu-Takayanagi surface $\Gamma_{BC}$, because in this way we get a lower entanglement entropy. This point is somewhat implicit in \cite{takayanagi,phuc},\footnote{In particular, in our AdS-Rindler setting, it is consistent with the fact that the optimal purification is \emph{not} the familiar thermofield double \cite{phuc}.} and has been emphasized most clearly in \cite{takayanagi2,hty}. {}From this, we deduce that the difference between choosing $\Sigma$ as in Fig.~\ref{eopfig}b or a more generic extremal surface $\Sigma'$ as in Fig.~\ref{eopfig}c corresponds in the CFT to the choice of the optimal versus a suboptimal partition $A'=B'C'$. The crucial aspect here is that the optimization in (\ref{eopdef}) selects a specific purification (a choice of auxiliary degrees of freedom $A'$ and overall state $\ket{\psi}$), and once this is known, it is perfectly well-defined to consider the effect of using suboptimal partitions of $A'$. We are thus led to generalize (\ref{eopdef}) by using the optimal purification $(A',\ket{\psi})$ but prescribing a specific bipartition $A'=B'C'$:
\begin{equation}\label{eopdef2}
\mathrm{P}'(B:C|B')\equiv
S_{BB'}|_{(A',\ket{\psi})}~.
\end{equation}
For generic choices of $B'$, this yields \emph{an} entanglement of purification, as opposed to (\ref{eopdef}), which is \emph{the} entanglement of purification. We will refer to $\mathrm{P}'(B:C|B')$ as the entanglement of purification for the specified partition.

Based on the preceding discussion, if we define the holographic counterpart of (\ref{eopdef2}) as\footnote{For the assignment of the auxiliary degrees of freedom $A'$ to concrete locations on $\Gamma_A$, which would allow one to explicitly relate a given bipartition $B'C'$ to a specific $\Sigma'$, two recent developments that provide a one-to-one mapping between points on $A$ and on $\Gamma_A$ might prove useful. One is the `bit thread' picture of entanglement entropy, developed in \cite{bitthread,bitthread2}. The other is the recent observation \cite{wen} that bulk modular flow \cite{js,jlms,fl} induces such a mapping. For our purposes here, since we work purely on the gravity side, it is enough to know that each choice of $\Sigma'$ corresponds to \emph{some} bipartition of $A'$.}
\begin{equation}\label{eoph2}
P'(B:C|B')\equiv\frac{\cA(\Sigma')}{4G_N}~,
\end{equation}
then it is natural to conjecture that $P'=\mathrm{P}'$. This connection was hinted at in \cite{bao}. We have argued here that it is essentially a consequence of the conjecture that $P=\mathrm{P}$, although, strictly speaking, the implication runs in the opposite direction, because the latter identification is a special case of the former. Notice that the definitions (\ref{eopdef2}) and (\ref{eoph2}) are not limited to the case depicted in Fig.~\ref{eopfig}c, where $B$ and $C$ are contiguous, but make sense as well in the generic case depicted in Fig.~\ref{eopfig}a.

To summarize, \emph{we have found that curve segments $\tilde{\mathrm{C}}_n$ in the entanglement shade of our Rindler wedge $\cE_A$, which by definition cannot be associated with entanglement entropies in the CFT restricted to $A$, can be reconstructed using entanglements of purification.} In the process, one identifies an optimal purification $(A',\ket{\psi})$, under which $\cE_A$ is directly described as a pure state, with the auxiliary degrees of freedom $A'$ living on the Rindler horizon $\Gamma_A$. One then considers partitions $A'=B'C'$ that are generically suboptimal, and works with the associated geodesics $\Sigma'$. Ultimately, then, \emph{in the extended system $AA'$ these geodesics do allow us to encode the curve segments $\tilde{\mathrm{C}}_n$ in specific intervals\footnote{To avoid possible confusion, we emphasize that the prime in $I'(\lambda)$ is part of the name of the interval, and does not refer to differentiation.}  $I'(\lambda)=BB'$, and their lengths do encode entanglement entropies, $S_{I'(\lambda)}=S_{BB'}$.} In the next subsection, we will show how to use these lengths to define a differential version of the entanglement of purification, which will reproduce the lengths of the segments $\tilde{\mathrm{C}}_n$ that were a priori nonreconstructible.

Let us now explain how to deal with the case of curves that are not at constant Rindler time. Just like in the static case, such curves will have some number $\tilde{N}$ of segments $\tilde{\mathrm{C}}_n$ inside the entanglement shade. An important difference is that, for $u^{\bt}(\lambda)\equiv\bt'(\lambda)\neq 0$, the geodesics tangent to these segments will exit the wedge not through $\Gamma_A$, but through the null portions of $\p\cE_A$. A priori, such geodesics \emph{cannot} be associated with entanglements of purification. We can remedy this by using null alignment (\ref{nullalignment}), choosing $n^{\bt}(\lambda)=-u^{\bt}(\lambda)$ for all $\lambda$. This ensures that the geodesics shot along the reoriented vectors $U(\lambda)$ lie at constant Rindler time, and therefore exit $\cE_A$ through $\Gamma_A$, even if each such geodesic is at a different value of $\bt$. Notice that, unlike the situation we had in Secs.~\ref{shadesubsec} and \ref{mappingsubsec}, where $U$ was subject to the two constraints (\ref{ineqUvUtau}) and (\ref{ineqsmallUrho}) to directly achieve reconstructibility, the single requirement that we need here, $U^{\bt}=0$, can always be enforced, and determines $n^{\bt}(\lambda)$ uniquely (up to the choice of sign $\sigma$ in (\ref{ngeneral})). With this trick, then, it is straightforward to extend our use of entanglements of purification to the covariant case, proceeding exactly as in the $U^{\bt}=0$ case of entanglement entropies studied in \cite{poincarepoint}. This trick is the reason why in Section \ref{rindlergeneralsubsec} we did not need to work out the explicit form of time-dependent geodesics that cross the Rindler horizon.

It should be noted that the authors of \cite{nutsandbolts}, when discussing hole-ography for static curves in the global BTZ black hole \cite{btz}, had anticipated the need of resorting to geodesics that cross the horizon. They assumed that the information about their lengths would be available in the purification of the CFT state via its thermofield double, which is dual to the inclusion of a second asymptotic region for the black hole (or in our language, the complementary Rindler wedge seen in Fig.~\ref{3dfig}b). As we have seen, the recipe for full reconstruction of curves in AdS-Rindler has become much more explicit and compact here thanks to the use of null alignment \cite{hmw,poincarepoint} and entanglements of purification \cite{takayanagi,phuc}, concepts that were not available at the time of \cite{nutsandbolts}.

The power and generality of these concepts is such that our recipe for entanglement wedge reconstruction can be extended beyond the situation, analogous to \cite{nutsandbolts}, of static curves in a Rindler wedge. We have already explained that the recipe covers the case of time-dependent curves inside this wedge.   It also works for static curves in the most general entanglement wedge in AdS$_3$, arising from a region $A$ composed of an arbitrary number of disconnected subregions (the case of two subregions has been illustrated in Fig.~\ref{eopfig}a). Below (\ref{eopdef2}) we emphasized that our generalized version of entanglement of purification  makes sense in that setting too, so the curve segments $\tilde{\mathrm{C}}_n$ inside the entanglement shade will again be encoded in the non-optimal geodesics that we have denoted $\Sigma'$. This includes not just geodesics extending from the boundary to $\Gamma_A$, but also those that have both ends on $\Gamma_A$ (which do not exist when $A$ is connected).

 Moving beyond pure AdS, we will now argue that the recipe applies as well for a generic entanglement wedge $\cE_A$, arising from an arbitrary region $A$ in any state of a 2-dimensional field theory dual to a smooth bulk geometry $M$ in a theory with Einstein gravity.\footnote{Extensions beyond Einstein gravity would involve the generalizations of Ryu-Takayanagi \cite{rt} or Hubeny-Rangamani-Takayanagi \cite{hrt} developed in \cite{dong,camps,flm,bdhm}.} Here we will no longer assume that the geometry is static, so $\cE_A$ is now constructed with the Hubeny-Rangamani-Takayanagi \cite{hrt} surface associated with $A$, which we will still denote $\Gamma_A$. In this context, the time-dependent version \cite{takayanagi,phuc} of holographic entanglement of purification $P$ involves optimal geodesics $\Sigma$ ending on $\Gamma_A$. The variant that we have determined to be useful for curve reconstruction, $P'$, involves as before suboptimal geodesics $\Sigma'$ ending on $\Gamma_A$, possibly with one endpoint on the boundary.

 Consider first the case where $A$ is connected and $M$ is geodesically complete, which can be dealt with by an argument very similar to the one we gave four paragraphs above. As usual, a generic curve will have segments $\tilde{\mathrm{C}}_n$ inside the entanglement shade, which are associated with geodesics that are not fully contained within the wedge. Typically, these geodesics will exit the wedge through the null portions of $\p\cE_A$, perhaps just on one side, but possibly on both. We can use null alignment to reorient any such geodesic, imposing the single condition that \emph{one} of its endpoints hits $\Gamma_A$. We then have no room for maneuvering the other endpoint, so if it happens not to lie on the boundary or on $\Gamma_A$, we would a priori be uncertain about the interpretation, because such a geodesic would not be of the $\Sigma'$ type directly associated with an entanglement of purification $P'$. But  this cannot happen when $A$ is connected.
   The reason is the following. We can continue the geodesic beyond $\cE_A$, to find its final endpoints $p$ and $q$ on $\p M$. In the scenario that worries us, $p$ and $q$ are both outside of $\cD_A$. The fact that the geodesic touches $\Gamma_A$ guarantees that $p$ and $q$ are spacelike-separated from $A$, so we can choose a time slice in the field theory that contains all three of these objects. Under the stated assumption that $A$ is connected, the interval $\overline{pq}$ between $p$ and $q$ would then be contained inside $A^\mathsf{c}$, so by the nesting property of entanglement wedges \cite{densitymatrix,wall,nesting}, it would have to be the case that $\cE_{\overline{pq}}\subset\cE_{A^\mathsf{c}}$. But this contradicts the claim that the geodesic in question enters $\cE_A$. \emph{We conclude then that, even in this more general covariant setting, null alignment suffices to ensure that the segments $\tilde{\mathrm{C}}_n$ can always be reconstructed using entanglements of purification.}

 The final extension is to lift the requirement that $A$ be connected and $M$ be geodesically complete. In this case, we lose the possibility of bringing in the property of entanglement wedge nesting for the final part of the argument, so in general there will be geodesics needed for reconstruction that (even after their optimal reorientation via null alignment) have one endpoint on $\Gamma_A$ and the other on the null portion of $\p \cE_A$. We have noted above that this problem does not arise for static curves on pure AdS$_3$, and more generally, it is avoided for curves that happen to be located at a moment of time-reflection symmetry of an otherwise arbitrary geometry. Generally, though, we do need a field theory interpretation for geodesics exiting the wedge through the null portion of $\p \cE_A$.

 To relate such geodesics to an entanglement of purification, we must make an identification between points on $\Gamma_A$ and points on the rest of
 $\p \cE_A$, through some notion of time evolution for the purifying degrees of freedom $A'$. The natural notion is provided by bulk modular flow \cite{js,jlms,fl} (see in particular \cite{wen}, and also the previously mentioned `bit thread' picture \cite{bitthread,bitthread2}).
\emph{The key lesson here is that, once we learn from \cite{takayanagi,phuc,bao,takayanagi2,hty} that the purifying degrees of freedom $A'$ live on $\Gamma_A$, it is natural to conjecture that the optimal state $\ket{\psi}$ is dual to a spacetime geometry that is nothing more and nothing less than the entanglement wedge $\cE_A$. This provides a new, self-contained, instance of holographic duality, where modular evolution in the boundary theory is implemented by modular evolution in the bulk. In this context, all geodesics are available as ordinary entanglement entropies in the extended field theory that lives on $AA'$.} Strictly speaking, of course, there is no modular evolution for $\Gamma_A$ itself, so in practice one must take the degrees of freedom $A'$ to live on a regulated version $\Gamma_A$, akin to the stretched horizon familiar from discussions of black hole dynamics \cite{stretched}. This is directly analogous to what we do at the opposite side of the wedge, where we are accustomed to
associating the original degrees of freedom $A$ with a surface at some radial location $\br=\br_{\mathrm{max}}<\infty$ that serves as a UV cutoff.
A deeper investigation of this excised version of subregion duality would surely be worthwhile, but we leave it for future work.

 A different generalization involves the passage to $d$-dimensional field theories with $d>2$. Presumably, the story we have developed here can be so extended at least under the same special conditions that allow higher-dimensional discussions of differential entropy \cite{myers,cds,hmw}, but we will not pursue that direction here.

\subsection{Differential entropy and differential purification} \label{esubsec}

In the previous subsection we understood that, with the aid of entanglement of purification and null alignment,
all curve segments in the entanglement shade of an arbitrary entanglement wedge $\cE_A$ in any 3-dimensional bulk geometry
can be associated with families of intervals in the purified version of the boundary theory.  We will now carry out explicitly the final task for reconstruction, showing how to recover the lengths of generic curves using entanglements. For simplicity, we will work again with static curves in the AdS-Rindler wedge.

Consider first the original version of the boundary CFT$_2$, before we purify. Given an interval $I=(\bx_- ,\bx_+)$ at constant time, we know that the corresponding entanglement entropy, $S_I$, is determined by the length of the associated geodesic $\Gamma_I$, via the Ryu-Takayanagi formula (\ref{rt}). In Sec.~\ref{rindlerstaticsubsec}, we showed that this geodesic, expressed in terms of the endpoints of the interval, takes the form (\ref{staticgeodesicvpm}), i.e.,
\begin{equation}\label{rtI}
\br(\bx)=\frac{\cosh\left(\bx-\frac{\bx_++\bx_-}{2}\right)}{
\sqrt{\sinh^2\left(\frac{\bx_+-\bx_-}{2}\right)-\sinh^2\left(\bx-\frac{\bx_++\bx_-}{2}\right)}}~.
\end{equation}
The portion of this geodesic running from $\bx_i$ to $\bx_f$ is found to be
\be \label{staticlength}
\int\limits_{\bx_i}^{\bx_f} ds = \frac{L}{2} \log\Big( \frac{\bx-\bx_-}{\bx-\bx_+}\Big)\Big\lvert^{\bx_f}_{\bx_i}~.
\ee
As expected, this length diverges when $\bx_i\to\bx_-$ and/or $\bx_f\to\bx_+$, because we are then considering the entire geodesic, extending all the way to the boundary. To regulate this divergence, we introduce a radial UV cutoff at $\br=\br_{\mathrm{max}}\gg 1$. Through (\ref{rtI}), this is equivalent to performing the integral (\ref{staticlength}) only from $\bx_i=\bx_-+\epsilon$ to $\bx_f= \bx_+ - \epsilon$, where
\be
\epsilon = \frac{1}{2 \br_{\rm max}^2} \coth\Big( \frac{1}{2} (\bx_+ - \bx_-) \Big)~.
\ee
The entanglement entropy (reported for convenience in units of $4 G_N$) then acquires the form
\be\label{EE}
S(\bx_-,\bx_+) = L \ln \Big( 2 \br_{\mathrm{max}}^2  \frac{\sinh(\bx_+-\bx_-)}{\coth(\frac{1}{2}(\bx_+ - \bx_-))}\Big)~.
\ee
(Instead of using this bare quantity, one could choose to work with the holographically renormalized version of entanglement entropy, defined in \cite{marika}.)

In a similar fashion, for a geodesic of the type (\ref{staticgeodesicvhvinfty}), with one endpoint located at $\bx_{\mathrm{h}}$ on the Rindler horizon $\Gamma_A$ and the other at $\bx_\infty$ on the boundary, we can compute the UV-regulated length. As explained in the previous subsection, this is interpreted via (\ref{eoph2}) as the entanglement of purification \cite{takayanagi,phuc} between the two segments in the CFT demarcated by $\bx_{\infty}$, choosing a specific bipartition $B'C'$ for the purifying degrees of freedom $A'$, that corresponds to partitioning $\Gamma_{A}$ at $\bx_{\mathrm{h}}$.  The result (again in units of $4G_N$) is
\be
\label{EofP}
P'(\bx_\infty,\bx_{\mathrm{h}}) = L \ln \Big( \frac{1}{2 \br_{\rm max}\cosh(\bx_{\mathrm{h}} -\bx_{\infty})}\Big)~,
\ee
where again, $\br_{\mathrm{max}}$ denotes the UV cutoff. Put in other words, in the optimal purification of $\cE_A$, where the purifying degrees of freedom $A'$ are understood to live on $\Gamma_A$, the quantity $P'$ is simply the entanglement entropy $S_{I'}$ for the interval in the purified CFT that is dual to the interval $I'=(\bx_\infty,\bx_{\mathrm{h}})$ on $\p(\cE_A|_t)$.

Now, given an arbitrary static curve $\mathrm{C}$, we know from Section~\ref{mappingsubsec} that it consists of some number $N$ of segments $\mathrm{C}_n$ outside of the entanglement shade, and some number $\tilde{N}$ of segments $\tilde{\mathrm{C}}_n$ inside the shade. Each segment of the former type can be encoded in a family of geodesics with endpoints at the boundary, giving rise to a family of intervals $I(\lambda)=( \bx_-(\lambda), \bx_+(\lambda))$. We can combine the corresponding entanglement entropies $S_{I(\lambda)}$, given by (\ref{EE}), to form the differential entropy $E$ associated with the segment. This quantity was originally defined in \cite{hole-ography}, but the most compact and useful expression for it was written down in \cite{hmw}. Employing this formula and (\ref{EE}), we find\footnote{As explained in \cite{hmw}, the definition of $E$ can be given alternatively by differentiating with respect to $\lambda$ instead of $\overline{\lambda}$. Since the relation between the two definitions involves integration by parts, the boundary function $f_{E}(\lambda)$ identified in (\ref{fRindler}) would then be modified.}
\begin{equation}\label{diffEs}
E \equiv \int\limits_{\lambda_i}^{\lambda_f}  d\lambda~  \frac{\partial S(\bx_-(\lambda), \bx_+(\overline{\lambda}))}{\partial \overline{\lambda}} \Big\lvert_{\overline{\lambda}= \lambda}
=L \int\limits_{\lambda_i}^{\lambda_f} d\lambda~\coth(\tfrac{1}{2} (\bx_+ - \bx_-)) \bx_+'~.
\end{equation}

In order to get a concrete expression for $E$, we should substitute in (\ref{diffEs}) the values of $\bx_{\pm}(\lambda)$ in terms of the coordinates $(\bt,\bx(\lambda),\br(\lambda))$ of the bulk curve coordinates.
The association between the two arises from the fact that, for any given $\lambda$, the desired geodesic $\Gamma_{I(\lambda)}$ must pass through the given point on the curve, with the appropriate slope. {}From the solution (\ref{rhostat}), this means that the following two equations must be satisfied
\be
\br = \frac{\br_0 \cosh (\bx - \bx_0)}{\sqrt{1 - \br_0^2 \sinh^2 (\bx - \bx_0)}}~,
\ee
\be
\frac{\br'}{\bx'} = \frac{\br_0(1+\br_0^2) \sinh (\bx-\bx_0)}{(1-\br_0^2 \sinh^2(\bx-\bx_0))^{3/2}}~,
\ee
or in the opposite direction,
\be\label{xo}
\bx_0 = \bx - \arcsinh \Big( \frac{\br'/ \bx'}{\sqrt{\br^2(1+\br^2)^2 - \br'^2/\bx'^2}}\Big)~,
\ee
\be\label{ro}
\br_0 = \sqrt{\frac{\br^2 (1+ \br^2)^2 -  \br'^2/\bx'^2}{(1+\br^2)^2 + \br'^2/\bx'^2}}~.
\ee

%
%

Differentiating the entanglement entropy with respect to the parameter of the right endpoint, $\bx_+(\lambda)$, we substitute equations (\ref{xo}), (\ref{ro}) and (\ref{vpmstatic}) into the differential entropy expression (\ref{diffEs}), to obtain
\begin{align}\label{diffEs2}
E = L \int d\lambda~ & \Bigg[ \bx' \lvert \bx' \lvert \frac{(1+\br^2)^{3/2}}{\sqrt{(1+\br^2)^2 \bx'^2 + \br'^2}}
\\
&
+ \lvert \bx' \lvert  (1+\br^2)^{3/2}   \frac{ (1+\br^2) \br \br' \bx'' + \bx' \Big( (1+3\br^2)\br'^2 -(1+\br^2) \br \br'' \Big)}{\sqrt{(1+\br^2)^2 \bx'^2 + \br'^2} ((1+\br^2)^2\br^2 \bx'^2 - \br'^2) }
\nonumber
\\
&
-
(1+\br^2)^2 \bx' \br' \frac{  (1 + 2\br^2 + \br^4)\br \bx'^3 + (1+ \br^2) \br' \bx'' + (3 \br \br'^2 - (1 + \br^2 ) \br'') \bx'  }{ ((1 + \br^2)^2 \bx'^2 + \br'^2 )  ((1 + \br^2)\br^2 \bx'^2 - \br'^2)  }\Bigg] ~.
\nonumber
\end{align}


This expression looks quite different from the sought curve length,
\be\label{length}
\cA =  L \int\limits_{\lambda_i}^{\lambda_f} d\lambda \sqrt{(1+\br^2)\bx'^2 + \frac{\br'^2}{1+\br^2}}~.
\ee
However, we must recall that there is no reason for the integrands in these two formulas to match directly. In the context where differential entropy was originally defined and explored \cite{hole-ography,myers,hmw}, the curves under consideration were closed (or infinite with a periodicity condition at infinity), so the claim that $\cA=E$ requires only that the integrands in (\ref{diffEs2}) and (\ref{length}) differ at most by a total derivative. In the case of open curves, considered in \cite{nutsandbolts,poincarepoint} and needed for the segments $\mathrm{C}_n$ under consideration here, this total derivative, upon integration, will give rise to a boundary function, that we call $f_{E}(\lambda)$. Importantly, this boundary function itself can be interpreted in terms of entanglement entropy in the CFT \cite{nutsandbolts,poincarepoint}.

Because of the close analogy between the case of the Poincar\'e wedge studied in \cite{poincarepoint} and the AdS-Rindler wedge that we are considering here, we can anticipate the form of the boundary function $f_{E}$. For this purpose, consider the tangent geodesic,
$\Gamma(\lambda)$,
to the point $(\bx(\lambda),\br(\lambda))$ on the curve labeled by $\lambda$. Then, as an ansatz, we propose that $f_{E}$ will turn out to be given by the length of the arc of this geodesic that stretches from $(\bx(\lambda),\br(\lambda))$ to $(\bx_+^\epsilon,\br_{\rm max})$. Using the alternative parametrization (\ref{rhostat}), we find that this distance takes the form
\be\label{fRindler}
f_{E}(\lambda) \equiv \int\limits_{\bx}^{\bx_+^\epsilon}ds =L\Big[\log\Big(\frac{2 \br_{\rm max}}{\br_0} \Big) -  \arctanh \Big(\sqrt{1+\br_0^2} \tanh(\bx - \bx_0) \Big)\Big]~.
\ee
And indeed, by means of (\ref{xo}) and (\ref{ro}), it can be shown that (\ref{fRindler}) is precisely what we need to accomplish the desired equality between  (\ref{diffEs2}) and (\ref{length}), namely
\begin{equation}\label{AEf}
\cA = E - f_{E}(\lambda_f) + f_{E}(\lambda_i)~.
\end{equation}
For closed curves, the boundary contribution of course drops out.


Let us now move on to the more interesting case of a curve segment $\tilde{\mathrm{C}}_n$ inside the entanglement shade, where the criterion (\ref{criterium1}) is not obeyed, and we need to resort to geodesics with one end on the horizon. In order to deal with it in complete parallel with our preceding analysis, we propose the new notion of \emph{differential purification}, denoted by $D$. This quantity is constructed using the family of geodesics $\Gamma_{I'(\lambda)}$  associated with the intervals $I'(\lambda)=(\bx_{\infty}(\lambda), \bx_{\mathrm{h}}(\lambda))$ in the purified CFT. We imitate the procedure in (\ref{diffEs}), differentiating the entanglement of purification (\ref{EofP}) to obtain
\begin{equation}\label{diffP}
D \equiv  \int d\lambda~  \frac{\partial P'(\bx_\infty(\lambda), \bx_{\mathrm{h}}(\overline{\lambda}))}{\partial \overline{\lambda}} \Big\lvert_{\overline{\lambda}=\lambda}
= - L \int d\lambda~  \tanh(\bx_\infty - \bx_{\mathrm{h}}) \bx_{\mathrm{h}}'~.
\end{equation}
The main novelty in this expression is that we are varying the location of the point at the horizon, $\bx_{\mathrm{h}}(\lambda)$. The justification for this is
that, in bulk description of the optimally-purified CFT, the horizon $\Gamma_A$ is exactly on the same footing as the rest of the boundary of $\cE_A|_t$.\footnote{As explained in the previous footnote, we could alternatively vary the location of the boundary endpoint, $\bx_{\infty}(\lambda)$, and
the form of the boundary function $f_D(\lambda)$ would then differ from (\ref{fhorizon}).}

In this case, the equations analogous to (\ref{xo}) and (\ref{ro}) are
\be\label{xinfty}
\bx_\infty = \bx + {\rm arccsch}\Big(\frac{\sqrt{\br'^2 - (\br+\br^3)^2 \bx'}}{(1+\br^2)^{3/2}\bx'} \Big) - \arcsinh \Big(\frac{\br (1+\br^2) \bx'}{\sqrt{\br'^2 - (\br + \br^3)^2 \bx'^2}} \Big)~,
\ee

\be\label{xhor}
\bx_{\mathrm{h}} = \bx - \arcsinh \Big(\frac{\br (1+\br^2) \bx'}{\sqrt{\br`^2-(\br + \br^3)^2 \bx'^2}} \Big).
\ee

By substituting (\ref{xinfty}) and (\ref{xhor}) in (\ref{diffP}) we obtain
\be\label{diffPh}
D = L \int d\lambda \Big( \frac{\br (1+\br^2)^{3/2} \bx' ( (\br + 2\br^3 + \br^5) \bx'^3 + (\br' + \br^2 \br') \bx'' + \bx' (3\br \br'^2 - (1 + \br^2) \br'') ) }{( (\br + \br^3)^2 \bx'^2 - \br'^2 ) \sqrt{(1 + \br^2)^2 \bx'^2 + \br'^2}}\Big)~.
\ee
Motivated by our previous results for entanglement entropy and differential entropy in the Rindler wedge, we propose an expression for the boundary function: we expect $f(\lambda)$ to be the length of the geodesic tangent to the curve at the point $\lambda$, within the region $(\bx, \bx_{\mathrm{h}})$. The resulting expression is
\be \label{fhorizon}
f_{D}(\lambda) \equiv \int\limits_{\bx}^{\bx_{\mathrm{h}}}ds = - \frac{L}{2} \ln \Big(\frac{\sinh(\bx + \bx_\infty - 2\bx_{\mathrm{h}})}{\sinh(\bx_\infty - \bx)} \Big)~.
\ee
And indeed, we can verify that this is precisely what we need to attain the desired equality between the differential purification (\ref{diffPh}) 
and the length of the bulk curve (\ref{length}):
\begin{equation}\label{ADfh}
\cA = D - f_{D}(\lambda_f) + f_{D}(\lambda_i)~.
\end{equation}

We have thus succeeded in showing through explicit computation that curve segments inside the entanglement shade can be reconstructed using differential purification. Combined with the more familiar story of entanglement entropy, described before, this completes the demonstration of complete reconstructibility for arbitrary static curves in the AdS-Rindler wedge. {}From the arguments in the previous subsection we know that there is no obstruction for similarly reconstructing generic curves in
an arbitrary wedge $\cE_A$,
using again differential entropy and differential purification.

\section*{Acknowledgements}

It is a pleasure to thank C\'esar Ag\'on, Jan de Boer, Ben Freivogel, Daniel Olivas and Marika Taylor for useful discussions,
and Bartek Czech and Sagar Lokhande for valuable comments on the manuscript.
The work of RE and AG was partially supported by Mexico's National Council of Science and Technology (CONACyT) grant 238734 and DGAPA-UNAM grant IN107115. JFP was supported by the Netherlands
Organization for Scientific Research (NWO) under the VENI scheme.

\appendix

\section{Coordinate Transformations}\label{appendix}

We start with AdS$_3$ in global coordinates,
\begin{equation}\label{globalmetric}
 ds^2=\frac{L^2}{\cos^2\varrho}\left(-d\tau^2+d\varrho^2+\sin^2\!\varrho\,d\theta^2\right)~,
 \end{equation}
where $\tau\in(-\infty,\infty)$, $\varrho\in[0,\pi/2)$ and $\theta\in[0,2\pi)$. The transformation
 \begin{eqnarray}\label{globaltopoincare}
 t&=&\frac{L\sin\tau}{\cos\tau+\sin\varrho\cos\theta}~,\nonumber\\
 x&=&\frac{L\sin\theta\,\sin\varrho}{\cos\tau+\sin\varrho\cos\theta}~,\\
 z&=&\frac{L\cos\varrho}{\cos\tau+\sin\varrho\cos\theta}~,\nonumber
 \end{eqnarray}
brings the metric to
the familiar Poincar\'e form
\be\label{poincaremetric}
ds^2=\frac{L^2}{z^2}\left(-dt^2+dx^2+dz^2\right)=\frac{r^2}{L^2}\left(-dt^2+dx^2\right)+\frac{L^2}{r^2}dr^2~,
\ee
with $r=L^2/z$.
Next, we focus on a specific spatial region $A$ at constant $t$ in the boundary CFT: an interval of length $\ell$, which, without loss of generality, can be taken to be centered at $x=0$. The corresponding minimal surface $\Gamma_A$ is the semicircle $x^2+z^2=\ell^2$. The entanglement wedge of $A$, $\cE_A$, is a AdS-Rindler wedge, and we wish to transform to coordinates adapted to it.

In the CFT, the (inverse of the) conformal transformation
\begin{eqnarray}\label{conformal}
t&=&\frac{\ell\sinh\bt}{\cosh\bx+\cosh\bt}~,\nonumber\\
x&=&\frac{\ell\sinh\bx}{\cosh\bx+\cosh\bt}~,
\end{eqnarray}
maps the causal diamond $\cD_A$ to the full plane $\bt\in(-\infty,\infty)$, $\bx\in(-\infty,\infty)$. In so doing,
it transforms the reduced density matrix $\rho_A$ to a thermal density matrix, so the entanglement entropy $S_A$ reduces to thermal entropy \cite{ch}. The corresponding bulk transformation is \cite{chm}
\begin{eqnarray}\label{poincaretobtz}
t&=&\frac{\ell\sqrt{\mathfrak{r}^2-1}\sinh\bt}{\mathfrak{r}\cosh\bx+\sqrt{\mathfrak{r}^2-1}\cosh\bt}~,\nonumber\\
x&=&\frac{\ell\mathfrak{r}\sinh\bx}{\mathfrak{r}\cosh\bx+\sqrt{\mathfrak{r}^2-1}\cosh\bt}~,\\
z&=&\frac{\ell}{\mathfrak{r}\cosh\bx+\sqrt{\mathfrak{r}^2-1}\cosh\bt}~,\nonumber
\end{eqnarray}
where $\mathfrak{r}\in(1,\infty)$. Notice that $(\bt,\bx,\mathfrak{r})$ have been chosen to be dimensionless. In these coordinates, the bulk metric takes the planar BTZ form
\begin{equation}\label{btzmetric}
ds^2=L^2\left(-(\mathfrak{r}^2-1)d\bt^2+\mathfrak{r}^2d\bx^2+\frac{d\mathfrak{r}^2}{\mathfrak{r}^2-1}\right)\,.
\end{equation}
The presence of the horizon at $\mathfrak{r}=1$ encodes the thermal character of the state. This transmutation of what is originally an acceleration horizon in the CFT into the horizon of a bulk black hole was first examined in \cite{unruh}, in the context of the holographic implementation of the Unruh effect (where one is dealing with the special case where $A$ is semi-infinite).

The Rindler wedge $\cE_A$ is the exterior of the black hole, $\mathfrak{r}>1$. For our purposes it will be more intuitive to use the radial coordinate
\begin{equation}\label{btztorindler}
\br=\sqrt{\mathfrak{r}^2-1}~,
\end{equation}
which covers the entire range $\br\in(0,\infty)$, in direct analogy with the Poincar\'e wedge. Our final form for the metric is then
\begin{equation}\label{rindlermetric}
ds^2=L^2\left(-\br^2d\bt^2+(1+\br^2)d\bx^2+\frac{d\br^2}{1+\br^2}\right)~.
\end{equation}
The horizon is located at $\br=0$, and the boundary at $\br\to\infty$.

\end{document}